\documentclass[aip,jcp,reprint,noshowkeys,superscriptaddress]{revtex4-1}
\usepackage{bmpsize}
\usepackage{tabularx,float,bm,xcolor,microtype,multirow,amscd,amsmath,booktabs}
\usepackage{amssymb,amsfonts,physics,mhchem,upgreek,pifont,wrapfig,enumitem,siunitx,xspace,cancel}
\usepackage{chemfig}
\usepackage[normalem]{ulem}
\usepackage{tikz}
\usetikzlibrary{arrows}
\usetikzlibrary{external}
\usetikzlibrary{matrix}
\newcolumntype{Y}{>{\centering\arraybackslash}X}

\usepackage{hyperref}
\hypersetup{
    colorlinks=true,
    linkcolor=blue,
    filecolor=blue,      
    urlcolor=blue,   
    citecolor=blue
}

\DeclareSIUnit{\angstrom}{\mbox{\normalfont\AA}}

\UndeclareTextCommand{\l}{OT1}
\DeclareTextSymbolDefault{\l}{T1}

\makeatletter
\DeclareRobustCommand{\AA}{%
   \leavevmode
   \vbox{\ialign{##\cr
         \hidewidth\char'27 \hidewidth\cr
         \noalign{\nointerlineskip\kern-1.4ex}
         A\cr
   }}%
}
\makeatother

\usepackage{natbib}
  
\newcolumntype{d}[1]{D{.}{.}{#1}}

\definecolor{hughgreen}{HTML}{007525}

\newcommand{\SupMat}{\textcolor{blue}{supplementary material}\xspace}

\newcommand{\cA}{\mathcal{A}}

\newcommand{\cI}{\mathcal{I}}
\newcommand{\cJ}{\mathcal{J}}

\newcommand{\cT}{\mathcal{T}}

\newcommand{\bH}{\boldsymbol{H}}
\newcommand{\bO}{\boldsymbol{0}}

\newcommand{\bg}{\boldsymbol{g}}
\newcommand{\bQ}{\boldsymbol{Q}}

\newcommand{\bv}{\boldsymbol{v}}
\newcommand{\bs}{\boldsymbol{s}}

\newcommand{\bh}{\boldsymbol{h}}
\newcommand{\bc}{\boldsymbol{c}}
\newcommand{\bI}{\boldsymbol{1}}

\usepackage{stackengine}
\stackMath
\newcommand\tsup[2][2]{%
   \def\useanchorwidth{T}%
   \ifnum#1>1%
   \stackon[-1.2ex]{\tsup[\numexpr#1-1\relax]{#2}}{\mathchar"307E}%
   \else%
   \stackon[-1ex]{#2}{\mathchar"307E}%
   \fi%
}

\newcommand{\hH}{\hat{H}}

\newcommand{\Evar}{E_{\text{var}}}
\newcommand{\Eexact}{E_{\text{exact}}}
\newcommand{\Psivar}{\Psi_{\text{var}}}
\newcommand{\Psiext}{\Psi_{\cA}}
\newcommand{\Psiint}{\Psi_{\cI}}
\newcommand{\Ei}{E_{\cI}}
\newcommand{\Ea}{E_{\cA}}
\newcommand{\dE}{\delta E}
\newcommand{\Ept}{E_{\text{PT2}}}
\newcommand{\UCAM}{Yusuf Hamied Department of Chemistry, University of Cambridge, Lensfield Road, Cambridge, CB2 1EW, U.K.}

\newcommand{\LCPQ}{Laboratoire de Chimie et Physique Quantiques (UMR 5626), Universit\'e de Toulouse, CNRS, UPS, France.}

\begin{document}

\raggedbottom

\title{Rationale for the Extrapolation Procedure in Selected Configuration Interaction}
\author{Hugh~G.~A.~Burton}
\email{hgaburton@gmail.com}
\affiliation{\UCAM}
\author{Pierre-Fran\c{c}ois \surname{Loos}}
\email{loos@irsamc.ups-tlse.fr}
\affiliation{\LCPQ}
%\affiliation{\DOW}

%%%%%%%%%%%%%%%%%%%%%%%%%%%%%%%%%%%%%%%%%%%%%%%%%%%%%%
\begin{abstract}
Selected configuration interaction (SCI) methods have emerged as state-of-the-art methodologies for achieving high accuracy and generating benchmark reference data for ground and excited states in small molecular systems. However, their precision relies heavily on extrapolation procedures to produce a final estimate of the exact result. Using the structure of the exact electronic energy landscape, we provide a rationale for the common linear extrapolation of the variational energy as a function of the second-order perturbative correction. In particular, we demonstrate that the energy gap and the coupling between the so-called internal and external spaces are the key factors determining the rate at which the linear regime is reached. Starting from first principles, we also derive a new non-linear extrapolation formula that improves the post-processing of data generated from SCI methods and can be applied to both ground- and excited-state energies.
\end{abstract}
%%%%%%%%%%%%%%%%%%%%%%%%%%%%%%%%%%%%%%%%%%%%%%%%%%%%%%

%%%%%%%%%%%%%%%%%%%%%%%%%%%%%%%%%%%%%%%%%%%%%%%%%%%%%%
\maketitle
%%%%%%%%%%%%%%%%%%%%%%%%%%%%%%%%%%%%%%%%%%%%%%%%%%%%%%

%%%%%%%%%%%%%%%%%%%%%%%%%%%%%%%%%%%%%%%%%%%%%%%%%%%%%%
\section{Introduction}
%%%%%%%%%%%%%%%%%%%%%%%%%%%%%%%%%%%%%%%%%%%%%%%%%%%%%%
Selected configuration interaction (SCI), \cite{Bender1969,Whitten1969,Huron1973,Buenker1974} and related methods (such as density-matrix renormalisation group approaches \cite{White1992,White1993,Chan2011} and others \cite{Eriksen2017,Eriksen2018,Eriksen2019a,Eriksen2019b,Motta2018,Lee2020,Xu2018,Xu2020,Magoulas2021,Gururangan2021}), have taken a prominent role in modern electronic structure theory. \cite{Loos2020a,Eriksen2020,Eriksen2021} Their primary purpose is to calculate reference correlation and excitation energies in small molecular systems,\cite{Eriksen2020,Loos2020b,Caffarel2016b,Holmes2017,Chien2018,Loos2018,Loos2019,Loos2020c,Veril2021} for which they have demonstrated a remarkable ability to yield highly accurate estimates of full configuration interaction (FCI) results.
The numerous variations of SCI all perform a sparse exploration of the Hilbert space by selecting only the most energetically relevant determinants. This natural philosophy emerges from the observation that, among the incredibly large number of determinants in the FCI space, only a tiny fraction of them significantly contribute to the energy. 
{Modern versions of SCI include CIPSI (CI using a Perturbative Selection made Iteratively)\cite{Huron1973,Giner2013,Giner2015,Caffarel2016a,Garniron2017,Garniron2018,Garniron2019,Loos2020a,Loos2020b,Damour2021,Damour2023} adaptive sampling CI (ASCI), \cite{Schriber2016,Tubman2016,Tubman2018,Tubman2020} semistochastic heatbath CI (SHCI), \cite{Holmes2016,Holmes2017,Sharma2017,Chien2018,Yao2020,Yao2021,Larsson2022} and iterative CI (iCI). \cite{Liu2014,Liu2016,Lei2017,Zhang2020,Zhang2021} Stochastic CI methods, such as Monte Carlo CI (MCCI) \cite{Coe2018,Coe2022} and FCI quantum Monte Carlo (FCIQMC), \cite{Booth2009,Cleland2010,Blunt2015,Ghanem2019,Deustua2017,Deustua2018} follow a similar philosophy by using a stochastic representation to select the most 
important determinants.}

The SCI wave function corresponds to a truncated CI expansion constructed from determinants in some internal (or model) space $\cI$
\begin{equation}
\ket{\Psivar} = \sum_{I \in \cI} c_{I} \ket{I},
\label{eq:psivar}
\end{equation}
with the associated variational energy $\Evar = \mel{\Psivar}{\hH}{\Psivar}$, where we assume the normalisation of the variational wave function, i.e., $\braket{\Psivar}{\Psivar} = 1$. The accuracy of $\ket{\Psivar}$ can be assessed using the second-order Epstein--Nesbet perturbation correction, computed using the determinants $\{ \alpha \}$ that lie outside the model space (i.e.~in the external space $\cA$) as 
\begin{equation}
\Ept = - \sum_{\alpha \in \cA} 
 \frac{\abs*{\mel{\Psivar}{\hH}{\alpha}}^2}{H_{\alpha\alpha} - \Evar}, 
\label{eq:en2}
\end{equation}
where $H_{\alpha\alpha} = \mel*{\alpha}{\hH}{\alpha}$. {This perturbative approximation is traditionally derived using L\"owdin partitioning, as shown in Appendix~\ref{sec:LP}}. The exact FCI wave function and energy are indicated by the limit $\Ept \to 0^{-}$.

Despite the sparse exploration of the Hilbert space, these state-of-the-art methods still rely heavily on extrapolation procedures to produce final FCI estimates. \cite{Holmes2017,Garniron2019,Eriksen2020,Eriksen2021} In particular, {it is widely observed that an approximate linear relationship appears between $\Evar$ and $\Ept$ when $\Ept$ becomes small enough}, and thus a linear or quadratic extrapolation of $\Evar$ for $\Ept \to 0$ is generally used to estimate the exact energy for an unconverged SCI calculation. \cite{Holmes2017} The precision and reliability of this post-processing extrapolation procedure are critical in order to produce meaningful estimates. 
However, to the best of our knowledge, no theoretical justification for the linear (or otherwise) relationship between $\Evar$ and $\Ept$ has been proposed. 

%%%%%%%%%%%%%%%%%%
% figure 1
%%%%%%%%%%%%%%%%%%
\begin{figure}[htb]
\includegraphics[width=0.8\linewidth]{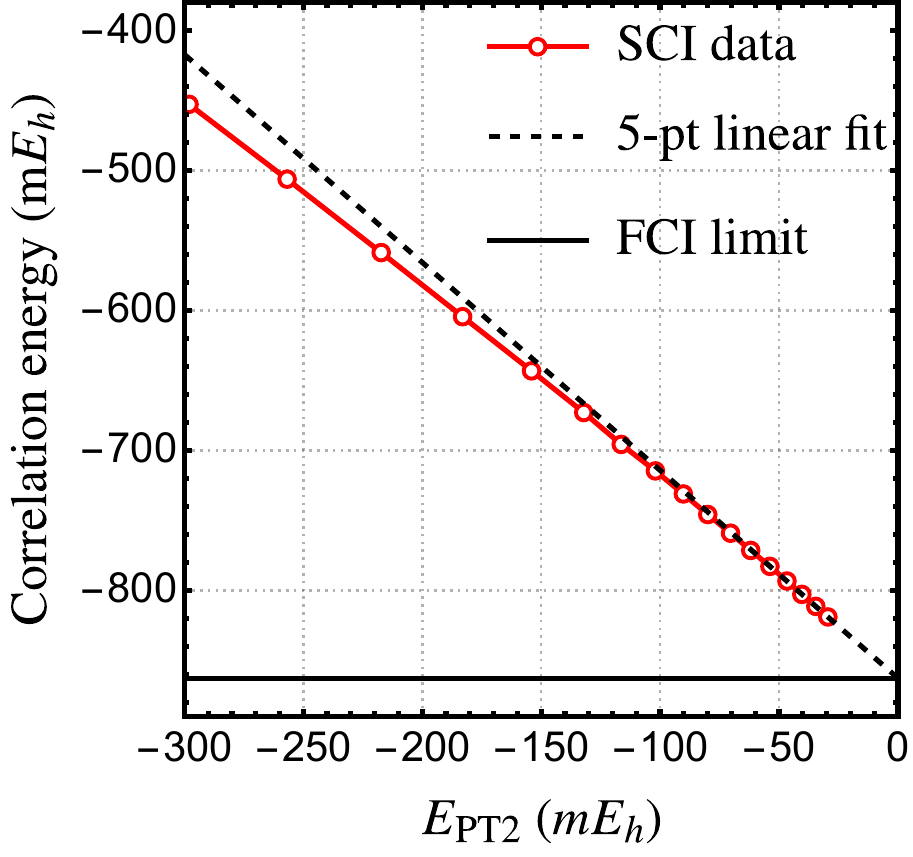}
\caption{Frozen-core (variational) correlation energy of benzene as a function of $\Ept$ computed in the cc-pVDZ basis, as described in Ref.~\onlinecite{Damour2021}.}
\label{fig:linearfit}
\end{figure}
%%%%%%%%%%%%%%%%%%

To illustrate this extrapolation procedure, Fig.~\ref{fig:linearfit} shows the evolution of the variational correlation energy of benzene as a function of $\Ept$, computed in the cc-pVDZ basis and within the frozen-core approximation. These SCI calculations were performed with \textsc{quantum package} using the CIPSI algorithm \cite{Garniron2019} and the data are extracted from Ref.~\onlinecite{Damour2021}. The FCI estimate of the correlation energy (solid black line in Fig.~\ref{fig:linearfit}) was estimated to be \SI{-862.890}{\milli\hartree} and was obtained by performing a five-point linear fit (dashed black line in Fig.~\ref{fig:linearfit}) of the CIPSI data. This estimate carries an error of the order of \SI{1}{\milli\hartree} and the fitting error was estimated to be \SI{0.266}{\milli\hartree}. From Fig.~\ref{fig:linearfit}, it is clear that, for sufficiently small $\Ept$, the variational quantity is linear with respect to $\Ept$. However, the SCI data deviate significantly from linearity for larger values of $\Ept$, which we shall address in detail later on.

In this Communication, we provide a rationale to justify the linear extrapolation of the (zeroth-order) variational energy as a function of the second-order perturbative energy. We adopt a geometric approach that considers the variational wave function as a point on the exact electronic energy landscape,\cite{Burton2021e} allowing the second-order perturbative correction to be derived from the local gradient and curvature of this energy landscape. Moreover, we investigate a two-state model in which an analytic relationship between $\Evar$ and $\Ept$ can be derived, leading to a novel parametrised non-linear formula that facilitates a more robust extrapolation procedure. 
%\PFL{Shall we squeeze a citation to \cite{Grimsley2022}?}

%%%%%%%%%%%%%%%%%%
\section{Rationale for the linear extrapolation}
%%%%%%%%%%%%%%%%%%

\begin{figure}
\includegraphics[width=0.9\linewidth]{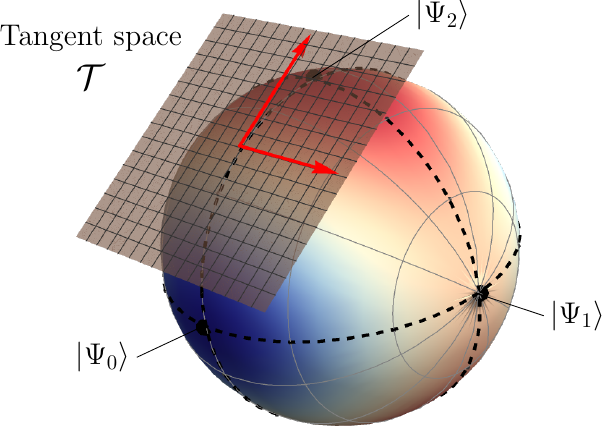}
\caption{Sketch of the exact electronic energy landscape in a three-dimensional Hilbert space. Eigenstates correspond to stationary points constrained to the surface of the unit sphere. At any point, the tangent space $\cT$ is spanned by the two vectors (red) that are orthogonal to the position vector.}
\label{fig:landscape}
\end{figure}

% LANDSCAPE PERSPECTIVE
The linear relationship between $\Evar$ and $\Ept$ can be derived from first principles by considering the structure of the electronic energy landscape. While Ref.~\onlinecite{Burton2021e} describes this energy landscape perspective in detail, the salient points are summarised here. Any normalised wave function in the full $N$-dimensional Hilbert space
\begin{equation}
\ket{\Psi} = \sum_{I=1}^{N} v_I \ket{I}
\end{equation}
can be represented by a vector $\bv$  subject to the normalisation constraint $\bv^\dagger \cdot \bv = 1$, which constrains the wave function to the surface of a hypersphere. The energy is given by the quadratic form 
\begin{equation}
E = \bv^\dagger \cdot \bH \cdot \bv
\end{equation} 
and exact eigenstates of the Hamiltonian correspond to stationary points of $E$ constrained to the surface of the hypersphere, as illustrated in Fig.~\ref{fig:landscape}. At any point on the hypersphere, the tangent space $\cT$ contains the vectors that are orthogonal to $\bv$, which can be collected as the columns of an $N \times (N-1)$ matrix $\bv_\perp$. These tangent vectors correspond to the states $\ket{T}$ that are orthogonal to $\ket{\Psi}$ and satisfy 
\begin{equation}
\hat{1} = \dyad{\Psi}{\Psi} + \sum_{T \in\cT} \dyad{T}{T},
\end{equation}
where $\hat{1}$ is the identity operator. A constrained step $\bs$ on this landscape is parametrised using a unitary transformation as
\begin{equation}
\ket*{\Psi(\bs)} = \exp(\sum_{T\in\cT} s_T \Big(\dyad{T}{\Psi} - \dyad{\Psi}{T} \Big) ) \ket{\Psi}.
\end{equation}
Assuming  real wave functions, the components of the constrained energy gradient are then
\begin{equation}
g_T = \eval{\pdv{E}{s_T}}_{\bs = \bO} 
%=
%\mel{\Psi_0}{[\hH,\hS]}{\Psi_0}
 = 2 \mel{T}{\hH}{\Psi},
\label{eq:grad}
\end{equation}
while the elements of the Hessian matrix of constrained second-derivatives become
\begin{equation}
\begin{split}
Q_{TT'} 
=
\eval{\pdv[2]{E}{s_T}{s_{T'}}}_{\bs = \bO}
= 2\mel{T}{\hH - E}{T'}.
\end{split}
\end{equation}

%%%%%%%%%%%%%%%%%%
% figure 2
%%%%%%%%%%%%%%%%%%
%%%%%%%%%%%%%%%%%%

So far, we have only considered the structure of the electronic energy landscape for an arbitrary wave function in the full Hilbert space. For a SCI variational wave function, the only non-zero elements of $\bv$ correspond to determinants included in the internal space, with coefficients $c_I$, as defined in Eq.~\eqref{eq:psivar}. The tangent vectors can then be split into two disjoint sets corresponding to the eigenstates within the internal space that are orthogonal to $\ket{\Psivar}$, denoted $\cJ = \{ \ket{J} \}_{J \neq \text{var}}$, and the determinants in the external space, giving $\cT = \cJ \cup \cA$, as illustrated in Fig.~\ref{fig:tangents}. {This separation of the internal and external spaces is further explained in Appendix~\ref{sec:LP}.} Since $\ket{\Psivar}$ is a CI solution within the internal space, such that $\mel{J}{\hH}{\Psivar}  = 0$, the gradient in Eq.~\eqref{eq:grad} is only non-zero in the direction of the tangent vectors in $\cA$.

%%%%%%%%%%%%%%%%%%
% figure 3
%%%%%%%%%%%%%%%%%%
\begin{figure}
\includegraphics[width=0.9\linewidth]{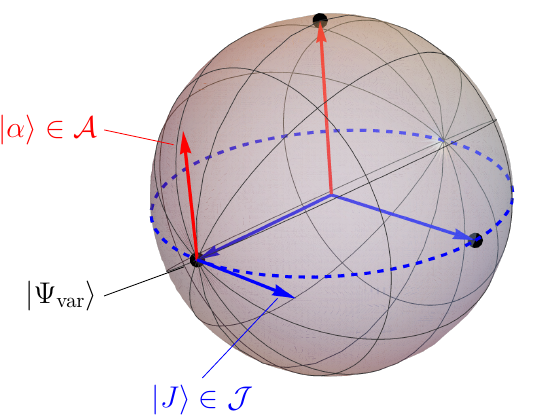}
\caption{Sketch of the tangent space construction $\cT = \cJ \cup \cA$. The variational space $\cI$ (dashed blue circle) is built from two configurations (blue vectors) and the external space $\cA$ contains one configuration (red vector). At $\ket{\Psivar}$, the tangent space is spanned by one tangent direction $\ket{J}$, which is locally parallel to the variational space $\cI$, and one orthogonal tangent direction in the external space $\ket{\alpha}$.}
\label{fig:tangents}
\end{figure}
%%%%%%%%%%%%%%%%%%

The local structure of the energy landscape around the variational wave function $\ket{\Psivar}$ is given by a second-order Taylor series expansion as
\begin{equation}
E(\bs) = \Evar + \bs^\dagger \cdot \bg + \frac{1}{2} \bs^{\dagger} \cdot \bQ \cdot \bs.
\end{equation}
Optimising this quadratic form with the Newton--Raphson step $\bs = -\bQ^{-1} \cdot \bg$ gives an estimate of the difference between the exact energy $\Eexact$ and $\Evar$ as
\begin{equation}
\Delta E = \Eexact - \Evar  \approx - \frac{1}{2} \bg^{\dagger} \cdot \bQ^{-1} \cdot \bg,
\label{eq:energyCorrection}
\end{equation}
or, equivalently,
\begin{equation}
\begin{split}
\Delta E \approx 
- 
\sum_{\alpha\alpha' \in \cA} \mel{\Psi}{\hH}{\alpha} \mel{\alpha}{(\hH - \Evar \hat{1})^{-1}}{\alpha'} \mel{\alpha'}{\hH}{\Psi},
\end{split}
\label{eq:de}
\end{equation}
where we have exploited the fact that the gradient is zero in the direction of the tangent vectors within $\cJ$. {Crucially,} the exact energy landscape {becomes} quadratic near an eigenstate.\cite{Burton2021e} {Therefore, Eq.~\eqref{eq:energyCorrection} becomes an  exact relationship when  $\ket{\Psivar}$ is sufficiently accurate, and we rigorously recover the linear asymptotic relationship }
\begin{equation}
{\Evar \sim \Eexact - \Delta E \qq{(as $\Delta E \to 0$).}}
\label{eq:de_relationship}
\end{equation}

{The steps that we have taken to derive this asymptotic relation are analytically rigorous. However,} the matrix elements $\mel{\alpha}{(\hH - \Evar\hat{1})^{-1}}{\alpha'}$  in Eq.~\eqref{eq:de} are too expensive to compute in practice. Instead, we can assume that the Hamiltonian is diagonally dominant in $\cA$ and take the leading order approximation
\begin{equation}
\hH^{(0)} = \sum_{II' \in \cI} \ket{I} H_{II'} \bra{I'} + \sum_{\alpha \in \cA} \ket{\alpha} H_{\alpha\alpha} \bra{\alpha},
\end{equation}
where $\hH^{(0)}$ is the zeroth-order Hamiltonian within the Epstein--Nesbet partitioning.
The energy correction then reduces to the second-order Epstein--Nesbet expression obtained in Eq.~\eqref{eq:en2} to give
\begin{equation}
\Delta E \approx -
  \sum_{\alpha \in \cA}  \frac{\abs*{\mel{\Psivar}{\hH}{\alpha}}^2}{H_{\alpha\alpha} - \Evar} = \Ept.
\label{eq:finalEpt}
\end{equation}
{Since $\Ept$ is the leading-order approximation to $\Delta E$, we obtain  $\Delta E \sim \Ept$ for $\Delta E \to 0$. Therefore, when $\ket{\Psi}$ is sufficiently close to an eigenstate, we obtain the asymptotic relationship 
\begin{equation}
\Evar \sim \Eexact - \Ept \qq{(as $\Ept \to 0$).}
\label{eq:ept_relationship}
\end{equation}
which justifies a linear extrapolation of $\Evar$ against $\Ept$.}

\begin{figure}[htb]
\includegraphics[width=0.8\linewidth]{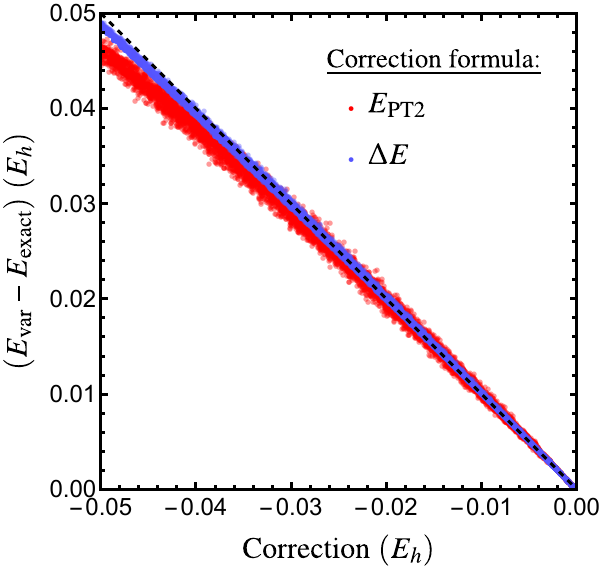}
\caption{{Numerical illustration of the asymptotic relationships, Eqs.~\eqref{eq:de_relationship} and \eqref{eq:ept_relationship}, for \ce{H2O} (STO-3G).
We randomly select 10,000 sets of determinants of varying sizes (including the HF ground-state determinant) to construct the SCI internal space and variational wave function.}}
\label{fig:scatter}
\end{figure}

{The only approximation employed in our derivation is to assume that Eq.~\eqref{eq:finalEpt} provides the dominant contribution to Eq.~\eqref{eq:de}.
From the exact energy landscape, we have shown that the asymptotic behaviour of $\Evar$ is linear for any SCI variational wave function as $\Ept \to 0$, and that the slope of this relationship is close to $-1$ for $\Evar \to \Eexact$. These properties are illustrated in Fig.~\ref{fig:scatter}, where we plot $\Evar$ against $\Delta E$ and $\Ept$ for 10,000 randomly selected SCI internal spaces in \ce{H2O} (STO-3G). We systematically target the ground state by ensuring that the HF determinant is always included in the internal space. The linear asymptote (black dashed) and the validity of $\Ept \sim \Delta E$ are clear for $\Evar \to \Eexact$.}

%Assuming that $\ket{\Psi}$ is sufficiently close to an eigenstate, such that the quadratic approximation to the exact energy landscape is valid, we can combine Eqs.~\eqref{eq:energyCorrection}, \eqref{eq:de}, and \eqref{eq:finalEpt} to obtain $\Evar \approx \Eexact - \Ept $, thus rationalising the linear extrapolation of $\Evar$ as a function of $\Ept$. 

%%%%%%%%%%%%%%%%%%
\section{Insights from a two-state model}
%%%%%%%%%%%%%%%%%%

In practice, linear or quadratic extrapolation procedures only work using a limited number of points {and} for well-converged calculations. As {illustrated in Fig.~\ref{fig:linearfit}}, the variational energy generally deviates away from linearity for larger values of $\Ept$. The cause of these deviations can be studied using a two-state model that represents the separation of the internal and external spaces in a SCI calculation, and for which the relationship between $\Evar$ and $\Ept$ can be analytically derived.

Our model contains individual states $\ket{\Psiint}$ and $\ket{\Psiext}$ representing wave functions in the internal and external spaces, respectively, with characteristic energies $\Ei$ and $\Ea$. The Hamiltonian matrix in this basis is then
\begin{equation}
\bH=
\begin{pmatrix}
\Ei & t
\\
t & \Ea
\end{pmatrix},
\end{equation}
where $t = \mel{\Psiint}{\hH}{\Psiext}$ represents the strength of the coupling between the internal and external spaces, and $\dE = \Ea - \Ei$ provides a measure of their energetic separation. The exact ground-state energy is
\begin{equation} 
\Eexact = \Ei + \frac{\dE}{2} - \sqrt{ \qty(\frac{\dE}{2})^2+ t^2}.
\label{eq:Eexact}
\end{equation}
The improvement of $\ket{\Psivar}$ during the course of a SCI calculation can be modelled by mixing $\ket{\Psiint}$ and $\ket{\Psiext}$ to give the parametrisation
\begin{equation}
\ket{\Psivar(\theta)} = \cos \theta \ket{\Psiint} +  \sin \theta \ket{\Psiext}, 
\end{equation}
with $0 \le \theta < 2\pi$. The corresponding energy is 
\begin{equation}
\Evar(\theta) = \Ei + \frac{\dE}{2}(1 - \cos 2 \theta) + t \sin 2\theta .
\label{eq:mod2var}
\end{equation}
Following a Taylor series expansion, the second-order correction is 
\begin{equation}
\Ept(\theta) = - \frac{1}{4} \frac{(2t \cos 2\theta  + \dE \sin 2 \theta)^2   }{  \dE \cos 2\theta - 2 t \sin 2\theta  }.
\label{eq:mod2pt}
\end{equation} 

By solving  Eq.~\eqref{eq:mod2pt} for $\theta$, we can invert these equations to express $\Evar$ in terms of $\Ept$ as
\begin{equation}
\Evar
=
\Ei + \frac{\dE}{2} -  \Ept
- \sqrt{ \qty(\frac{\dE}{2})^2 + t^2 + \qty(\Ept)^2}.
\label{eq:2linv}
\end{equation}
which naturally reduces to Eq.~\eqref{eq:Eexact} for $\Ept = 0$. This expression reveals that the more general form of $\Evar$ as a function of $\Ept$ involves a square-root term that deviates away from linearity, and that this departure from the linear regime is directly related to the energetic separation ($\dE$) and coupling strength ($t$) between the internal and external spaces. For $(\Ept)^2 \ll  \qty(\frac{\delta E}{2})^2 + t^2 $, we recover the linear behaviour $\Evar \approx \Eexact - \Ept$. In a real SCI calculation, we expect $\abs{t} \ll \abs{\frac{\delta E}{2}}$. Therefore, the larger the energy separation, the sooner the linear regime is reached, while a large coupling between $\cI$ and $\cA$ also leads more rapidly to the linear regime. These features are further demonstrated through the series expansion of $\Evar$ at small $\Ept$,
\begin{equation}
	\Evar = \Eexact - \Ept - \frac{\Ept^2}{2 \sqrt{ \qty(\frac{\dE}{2})^2 + t^2}} + \order{\Ept^3},
\end{equation}
which shows that the quadratic behaviour is minimal when $\abs{\dE}$ or $\abs{t}$ is large.
 
\begin{figure}[hb]
\includegraphics[width=0.8\linewidth]{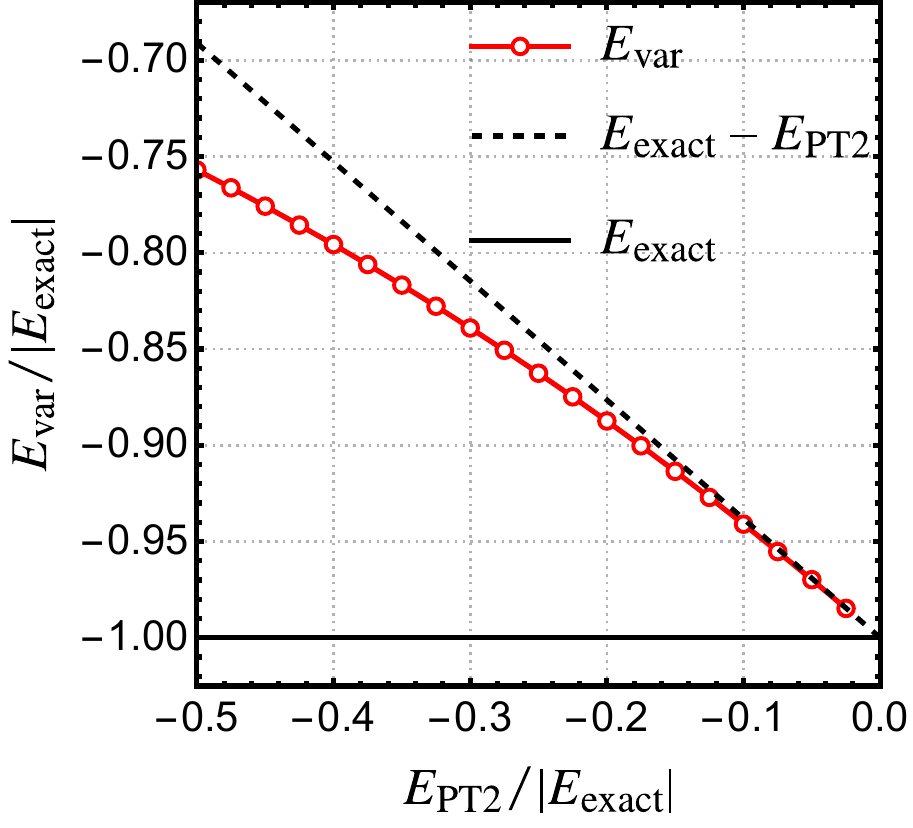}
\caption{$\Evar$ (red markers) as a function of $\Ept$ for the two-state system with $\Ei = -1$, $\delta E = 1$ and $t = 1$. These data deviate from the linear approximation (black dashed line) due to the square-root term in Eq.~\eqref{eq:2linv}.}
\label{fig:twolevel}
\end{figure}

This functional relationship can be illustrated by evaluating $\Evar$ for various values of $\Ept$ in the limit of interest, as shown in Fig.~\ref{fig:twolevel} for $\Ei = -1$, $\delta E = 1$ and $t = 1$. As $\Ept$ gets larger, $\Evar$ strongly deviates from linearity and bears a close similarity to real SCI data. \cite{Holmes2017,Chien2018,Garniron2019,Eriksen2020,Loos2020b,Loos2018,Loos2019,Loos2020c} Figure \ref{fig:linearfit} nicely illustrates this square-root behaviour for a realistic system, and the similarities between Figs.~\ref{fig:linearfit} and \ref{fig:twolevel} are striking.
{For realistic systems, the two-state model can be approximately engineered using a suitable transformation of the external space $\cA$ such that only 
one external state couples to the variational wave function through the Hamiltonian, as described in Appendix~\ref{sec:model_map}.}

%%%%%%%%%%%%%%%%%%
% figure 4
%%%%%%%%%%%%%%%%%%
%%%%%%%%%%%%%%%%%%

%%%%%%%%%%%%%%%%%%%%%%%%%%%%%%%%%%%%%%%%%%%%
\section{A non-linear extrapolation formula}
%%%%%%%%%%%%%%%%%%%%%%%%%%%%%%%%%%%%%%%%%%%%
The insights from our two-state model suggest that a non-linear functional form {may be} more suitable for extrapolating SCI data.
The separation of the model and perturbation space in SCI means that the relationship between $\Evar$ and $\Ept$ is not as straightforward as the model system.
However, as more determinants are added to the model space, the variational wave function follows a path towards the exact ground state that is likely to resemble Eq.~\eqref{eq:2linv}.
The concave form of Eq.~\eqref{eq:2linv} suggests that a linear extrapolation procedure will generally underestimate the exact correlation energy 
{and that any two-point linear fit would provide an upper bound to the exact energy.}
Therefore, we propose a new non-linear extrapolation formula
\begin{equation}
\Evar(\Ept; a, b, c) = a + \frac{\abs{c}}{2} - b \Ept -  \sqrt{\qty(\frac{c}{2})^2 + (b \Ept)^2},
\label{eq:nonlinear}
\end{equation}
where $a$, $b$, and $c$ are fitting parameters. This expression corresponds to a form of quadratic approximant, which has been previously used in the resummation 
of divergent perturbation expansions. \cite{Marie2021,Goodson2012,Goodson2019,Mayer1985} Crucially, Eq.~\eqref{eq:nonlinear} reduces to a linear fit for $\abs{\Ept} \ll \abs{\frac{c}{2b}}$ and can reproduce the observed non-linearity for larger $\Ept$. The fitted value of $a$ provides the estimate for the FCI result.

%%%%%%%%%%%%%%%%%%
% figure 5
%%%%%%%%%%%%%%%%%%
\begin{figure}[b]
\includegraphics[width=0.8\linewidth]{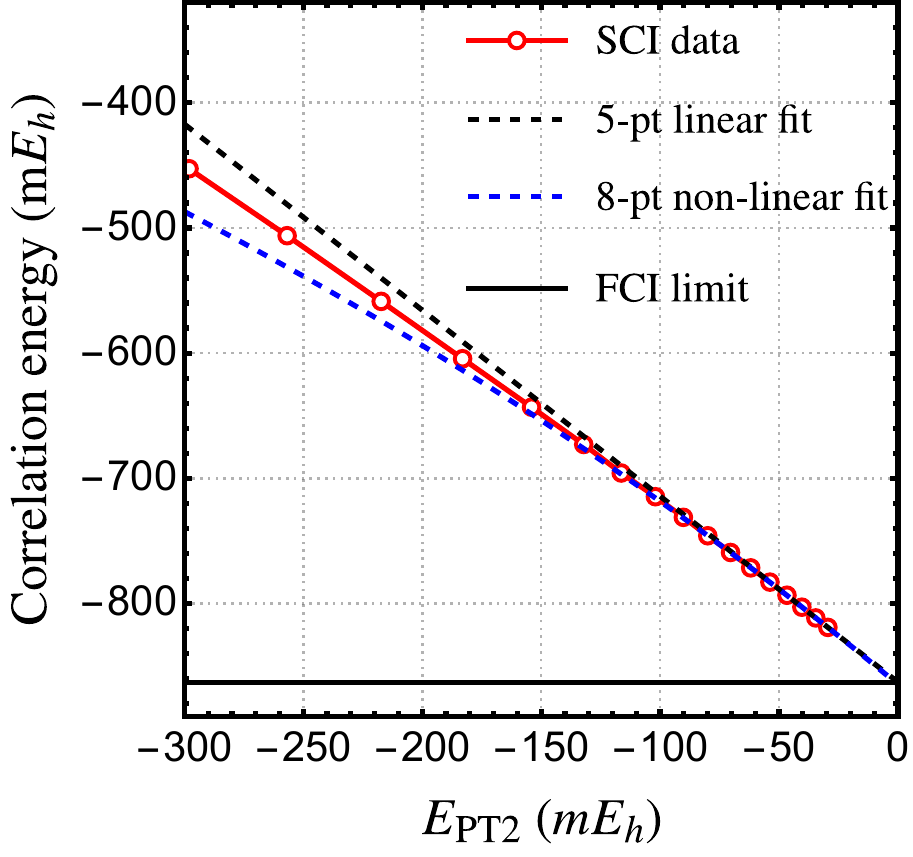}
\caption{Comparison of the linear and non-linear extrapolation procedure for the correlation energy of benzene computed in the cc-pVDZ basis, using the SCI data from Ref.~\onlinecite{Damour2021}.}
\label{fig:nlfit}
\end{figure}
%%%%%%%%%%%%%%%%%%

%%%%%%%%%%%%%%%%%%
% figure 6
%%%%%%%%%%%%%%%%%%
\begin{figure*}
\includegraphics[width=0.49\linewidth]{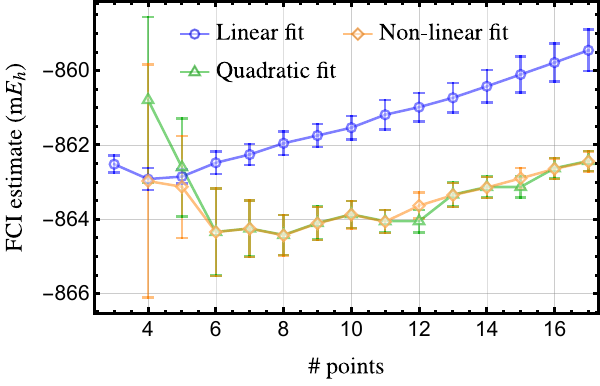}
\includegraphics[width=0.49\linewidth]{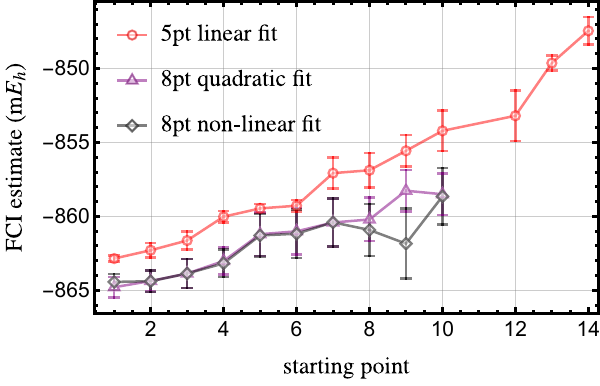}
\caption{Evolution of the FCI estimate of the correlation energy of benzene computed in the cc-pVDZ basis. \textit{Left}: A variable number of fitting points is considered in the {linear (blue), quadratic (green), or non-linear (orange)} fits, with each fit starting from the point associated with the smallest value of $\Ept$. \textit{Right}: A fixed-length window of consecutive points is used for the {linear (red), quadratic (purple), or non-linear (black)} fits. The point of index 1 corresponds to the smallest value of $\Ept$, and a higher index for the starting point corresponds to a larger $\Ept$ correction. The error bars indicate the standard errors associated with the fitting procedure.}
\label{fig:fit_benzene}
\end{figure*}
%%%%%%%%%%%%%%%%%%

The variation of the FCI estimate of the correlation energy of benzene computed in the cc-pVDZ basis using a linear or non-linear extrapolation procedure is compared in Fig.~\ref{fig:nlfit}. These data show that the non-linear formula can accurately fit the SCI data at larger $\Ept$ values than the linear procedure. In the left panel of Fig.~\ref{fig:fit_benzene}, we study the influence of the number of points included in the {linear, quadratic, or non-linear fits}, each of them starting from the point associated with the smallest value of $\Ept$. The error bars indicate the standard errors associated with the fitting procedure. 
{The fits are weighted according to the inverse square of the perturbative corrections.}
Although the FCI estimates obtained from the non-linear formula [see Eq.~\eqref{eq:nonlinear}] have larger fitting errors for a small number of points, they quickly stabilise and remain relatively consistent compared to those obtained from the linear procedure, which rises much faster. Consequently, a larger number of points can, and should, be employed for the non-linear extrapolation formula, while the linear extrapolation procedure becomes systematically worse when more points are used.
{Our results also demonstrate that, except for small numbers of points, the quadratic and non-linear fits are nearly identical. However, in contrast to the quadratic fit, the non-linear formula will never predict a maximum in $\Evar$ vs $\Ept$.}

In the right panel of Fig.~\ref{fig:fit_benzene}, we consider a fixed-length window of consecutive points for the {linear, quadratic, and non-linear fits}, but we vary the index of the starting point, with the point at index 1 corresponding to the smallest value of $\Ept$ (i.e., the higher the index of the starting point, the larger the $\Ept$ correction). 
{Using 5 points for the linear fit and 8 points for the quadratic and non-linear fits}, we again show that {the quadratic and non-linear procedures are slightly more stable than the linear version} with respect to the starting point, although {all} extrapolation schemes eventually underestimate the correlation energy for larger $\Ept$ values.
These results indicate that, compared to the linear approach, the non-linear extrapolation procedure can provide a more accurate estimate of the FCI result for SCI calculations that are less well converged. 
%\PFL{Maybe adding a comment about the quadratic fit here.}

%%%%%%%%%%%%%%%%%%
% figure 7
%%%%%%%%%%%%%%%%%%
\begin{figure*}
\includegraphics[height=0.34\linewidth,trim=0pt -7pt 0pt 0pt, clip]{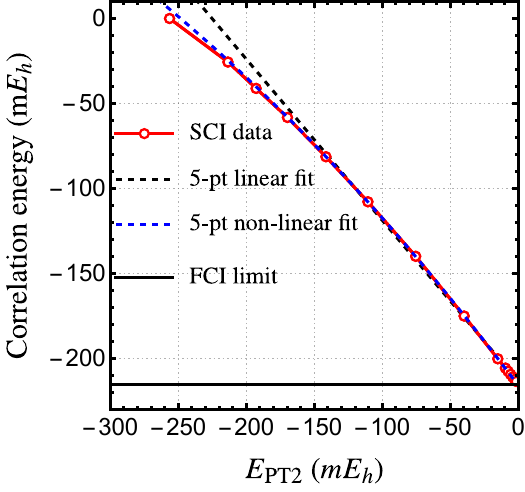}
\hspace{1em}
\includegraphics[height=0.34\linewidth]{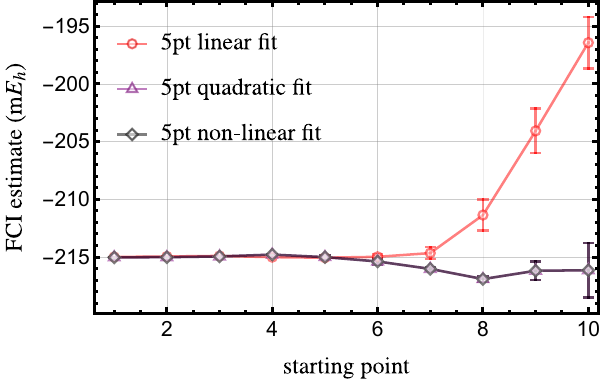}
\caption{{\textit{Left}: Comparison of the linear and non-linear extrapolation procedure for the correlation energy of water computed in the cc-pVDZ basis. The linear and non-linear weighted fits are performed with 5 points with minimal and maximal values of $\Ept = \SI{9.5d-3}{\hartree}$ and $\Ept = \SI{1.1d-1}{\hartree}$, respectively. \textit{Right}: A 5-point window of consecutive points is used for the linear (red), quadratic (purple), or non-linear (black) fits. The point of index 1 corresponds to a value of $\Ept = \SI{1.6d-3}{\hartree}$, and a higher starting point index corresponds to a larger $\Ept$ correction. The error bars indicate the standard errors associated with the fitting procedure.}}
\label{fig:fit_water}
\end{figure*}
%%%%%%%%%%%%%%%%%%

{The benzene example offers a practical illustration that is close to the upper limit of what is currently achievable using SCI methods. \cite{Eriksen2020,Loos2020b,Damour2021} However, since the exact FCI correlation energy remains unknown, there are potential ambiguities in the comparison of different extrapolation procedures.
Instead, we turn to another example where the Hilbert space is significantly smaller, allowing the SCI calculations to be converged to the FCI limit. We consider the water molecule with the geometry used in Ref.~\onlinecite{Loos2018}, for which the exact (frozen-core) correlation energy in the cc-pVDZ basis set is \SI{-215.027}{\milli\hartree}. We conduct a ground-state CIPSI calculation based on HF orbitals. The calculation is intentionally halted prematurely to assess the feasibility of estimating the true correlation energy with smaller variational spaces. Once again, we evaluate and compare the performance of linear, quadratic, and non-linear fits.}

{The left panel of Fig.~\ref{fig:fit_water} reports the convergence of the variational correlation energy as a function of $\Ept$ as well as the 5-point linear and non-linear fits obtained with minimal and maximal values of $\Ept = \SI{9.5d-3}{\hartree}$ and $\Ept = \SI{1.1d-1}{\hartree}$, respectively. (The quadratic fit is almost indiscernible from the non-linear fit.) The right panel of Fig.~\ref{fig:fit_water} shows the FCI estimates produced from the linear, quadratic, and non-linear fits relying on a 5-point window of consecutive points. (The point of index 1 corresponds to a value of $\Ept = \SI{1.6d-3}{\milli\hartree}$.) Once again, the quadratic and non-linear fitting procedures are nearly identical and have considerably greater stability for less well-converged calculations compared to the linear version. For example, considering a set of only 351 determinants as the largest variational space (which is associated with a PT2 value of \SI{7.5d-2}{\hartree}), the estimated FCI value is \SI{-216.192}{\milli\hartree} which is in error by around \SI{1}{\milli\hartree}, while the linear fit predicts \SI{-204.07}{\milli\hartree}.}

%%%%%%%%%%%%%%%%%%
\section{Concluding Remarks}
%%%%%%%%%%%%%%%%%%
In this Communication, we have proposed a theoretical rationale for the linear extrapolation procedure of the variational zeroth-order energy, $\Evar$, as a function of the second-order perturbative correction, $\Ept$, commonly employed in SCI methods. Our derivation is based on connecting the SCI variational wave function to the properties of the underlying electronic energy landscape. The accuracy of the extrapolation of $\Evar$ for $\Ept \to 0$ is critical for accurately estimating the final FCI value. 

Our investigations led us to the discovery of a novel non-linear extrapolation formula that more effectively captures the behaviour of $\Evar$ for larger $\Ept$ values, thereby enhancing the robustness of extrapolations toward the FCI limit. Based on a two-state model, we derived the analytic form of this non-linear extrapolation formula and examined its mathematical properties. Specifically, we illustrated that the rate at which the linear regime is attained is primarily determined by the energetic gap and the coupling between the internal and external spaces. As a concrete example, we studied the ground-state correlation energy of benzene {and water}, which illustrates the versatility and reliability of this new extrapolation procedure.

{We anticipate that our findings will provide a useful contribution to the computational toolbox for performing SCI calculations. In particular, introducing a mathematically motivated non-linear extrapolation alongside linear and quadratic approaches provides an alternative route to gain confidence in the validity of extrapolation procedures. While our numerical results indicate that the non-linear extrapolation performs similarly to a quadratic fit, the non-linear approach has the advantage that it cannot create a maximum in $\Evar$ vs $\Ept$, preventing catastrophically unphysical extrapolations.}
%{We anticipate that this study will facilitate SCI calculations on larger Hilbert spaces, while providing confidence in the validity of extrapolation procedures.} \PFL{Shall we modify this sentence?}
Furthermore, we expect that our two-state model and our framing of the SCI approach within the electronic energy landscape framework will allow  
other intriguing mathematical aspects of these methods to be explored in the future. 

%%%%%%%%%%%%%%%%%%%%%%%%%%%%%%%%%%%%%%%%%%%%%%%%%%%%%%%%%%%%%%
\section*{Acknowledgements}
\label{sec:Acknowledgements}
%%%%%%%%%%%%%%%%%%%%%%%%%%%%%%%%%%%%%%%%%%%%%%%%%%%%%%%%%%%%%%
HGAB thanks Downing College, Cambridge for support through the Kim and Julianna Silverman Research Fellowship. PFL thanks financial support from the European Research Council (ERC) under the European Union's Horizon 2020 research and innovation programme (Grant agreement No.~863481).

%%%%%%%%%%%%%%%%%%%%%%%%%%%%%%%%%
\section*{Supplementary Material}
%%%%%%%%%%%%%%%%%%%%%%%%%%%%%%%%%
See the \SupMat for the SCI data of Fig.~\ref{fig:linearfit} and the raw data associated with {Figs.~\ref{fig:fit_benzene} and \ref{fig:fit_water}}.

%%%%%%%%%%%%%%%%%%%%%%%%%%%%%%%%%%%%%%
\section*{Data availability statement}
%%%%%%%%%%%%%%%%%%%%%%%%%%%%%%%%%%%%%%
The data that supports the findings of this study are available within the article and its supplementary material.

\appendix
%%%%%%%%%%%%%%%%%%%%%%%%%%%%%%%%%
\section{Derivation of the Epstein--Nesbet correction}
\label{sec:LP}
%%%%%%%%%%%%%%%%%%%%%%%%%%%%%%%%%
{The formal connections between internal and external spaces can be understood through the structure of the Hamiltonian matrix in the union space $\cI \cup \cA$. \cite{Garniron2018} Considering these two disjoint subspaces, the corresponding Schr\"odinger equation reads
\begin{equation} \label{eq:Schrodinger_block}
	\mqty( \bH_{\cI} & \bh^\dag \\ \bh & \bH_{\cA} )
	\cdot
	\mqty( \bc_{\cI} \\ \bc_{\cA} ) 
	= E 
	\mqty( \bc_{\cI} \\ \bc_{\cA} ) ,
\end{equation}
where 
\begin{subequations}
\begin{align}
	\bH_{\cI} & = \sum_{II' \in \cI} \ket{I} H_{II'} \bra{I'},
	\\
	\bH_{\cA} & = \sum_{\alpha\alpha' \in \cA} \ket{\alpha} H_{\alpha\alpha'} \bra{\alpha'}
\end{align}
\end{subequations}
are the Hamiltonian blocks associated with the internal and external spaces, respectively, and $\bh$ is the corresponding coupling block with elements $\mel{\alpha}{\hH}{I}$.
Using the L\"owdin partitioning technique, \cite{Lowdin1951} the second row of Eq.~\eqref{eq:Schrodinger_block} yields
\begin{equation} \label{eq:2nd_row}
	\bc_{\cA} = \qty( E \bI - \bH_{\cA} )^{-1} \cdot \bh \cdot \bc_{\cI},
\end{equation}
while, from the first row, one gets
\begin{equation} \label{eq:1st_row}
	\bH_{\cI} \cdot \bc_{\cI} + \bh^\dag \cdot \bc_{\cA} = E \bc_{\cI}.
\end{equation}
Substituting Eq.~\eqref{eq:2nd_row} into Eq.~\eqref{eq:1st_row} leads to the non-linear Schr\"odinger equation
\begin{equation}
	\bH^\text{eff}_{\cI}(E) \cdot \bc_{\cI} = E \bc_{\cI},
\end{equation}
%or
%\begin{equation}
%	E = \bc_{\cI}^\dag \cdot \bH^\text{eff}_{\cI}(E) \cdot \bc_{\cI},
%\end{equation}
where the energy-dependent effective Hamiltonian reads \cite{Malrieu1985}
\begin{equation} \label{eq:Heff}
	\bH^\text{eff}_{\cI}(E) = \bH_{\cI} + \bh^\dag \cdot \qty( E \bI - \bH_{\cA} )^{-1} \cdot \bh.
\end{equation}
solving this non-linear eigensystem is hard and unpractical due, mainly, to its energy dependence and the inversion of a large matrix.
Therefore, one usually relies on educated approximations by i) setting $E \approx E_{\cI} = \bc_{\cI}^\dag \cdot \bH_{\cI} \cdot \bc_{\cI}$ in the right-hand side of Eq.~\eqref{eq:Heff} in order to wash away the energy dependence, hence the non-linearity, and ii) enforcing a diagonal approximation for $\bH_{\cA}$, that is
\begin{equation}
	\bH_{\cA} \approx \sum_{\alpha \in \cA} \ket{\alpha} H_{\alpha\alpha} \bra{\alpha}.
\end{equation}
One then ends up with the  approximate energy expression
\begin{equation}
	E \approx E_{\cI} + \sum_{\alpha \in \cA} \frac{\abs*{\mel{\alpha}{\hH}{I}}^2}{E_{\cI} - H_{\alpha\alpha}},
\end{equation}
which is equivalent to the second-order Epstein-Nesbet perturbation correction in Eq.~\eqref{eq:en2}.}

%%%%%%%%%%%%%%%%%%%%%%%%%%%%%%%%%
\section{Connecting the two-state model to realistic systems}
\label{sec:model_map}
%%%%%%%%%%%%%%%%%%%%%%%%%%%%%%%%%
{For realistic systems, the two-state model can be approximately engineered using a singular value decomposition (SVD) of the gradient vector. Since the relevant components of the gradient [see Eq.~\eqref{eq:grad}] form a $N_\cA \times 1$ vector (where $N_\cA$ is the number of external states), an SVD can be used to transform the external space $\cA$ such that only one direction has a non-zero gradient, meaning that only one state $\ket{\Psiext} = \sum_{\alpha \in \cA} c_\alpha \ket{\alpha}$ couples to the variational wave function $\ket{\Psiint}$ through the Hamiltonian. Under this transformation, we obtain $ \Ei = \mel{\Psiint}{\hH}{\Psiint}$, $t = \mel{\Psiext}{\hH}{\Psiint}$ and $\Ea = \mel{\Psiext}{\hH}{\Psiext}$, where remaining indirect  Hamiltonian coupling terms are ignored.
This process can be illustrated for a Hamiltonian with two states in both the internal and external spaces, giving
\begin{equation}
\bH = \begin{pmatrix}
H_{00} & 0 &  H_{0\alpha} & H_{0 \beta}  \\
0 & H_{11} &   H_{1\alpha} & H_{1 \beta} \\
H_{\alpha 0} &H_{\alpha 1}     &  H_{\alpha \alpha} & H_{\alpha \beta} \\
H_{\beta 0} & H_{\beta 1}     &  H_{\beta \alpha} & H_{\beta \beta} 
\end{pmatrix}.
\end{equation}
The $H_{\alpha0}=\mel{\alpha}{\hH}{0}$ elements directly couple the internal and external spaces and are  proportional to the relevant gradient components $g_\alpha = 2H_{\alpha0}$. The SVD ensures that $\ket{0}$ directly couples to only one external state, giving
\begin{equation}
\bH = \begin{pmatrix}
H_{00} & 0 &  H_{1\alpha'}  & 0  \\
0 & H_{11} &   H_{1\alpha'} & H_{1 \beta'} \\
H_{\alpha'0} & H_{\alpha' 1}     &  H_{\alpha' \alpha'} & H_{\alpha' \beta'} \\
 0 & H_{\beta' 1}     &  H_{\beta' \alpha'} & H_{\beta' \beta'} \\
\end{pmatrix}
\end{equation}
Ignoring the indirect coupling elements $H_{\alpha' 1}$ and $H_{\alpha' \beta'}$ then gives the two-state model with $\Ei = H_{00}$, $t=H_{\alpha'0}$, and $\Ea = H_{\alpha' \alpha'}$.
Furthermore, since the  states within $\cI$ are decoupled by solving the CI problem, we can construct this two-state model for any variational state by choosing $\ket{\Psiint}$ and performing the SVD on the corresponding gradient vector. Therefore, this two-state model can represent both ground and low-lying excited states using different values of $\Ei$, $\dE$, and $t$. While this transformation is not feasible in practice, it conceptually links  real SCI data and the present two-state model. }

%%%%%%%%%%%%%%%%%%%%%%%%%%%%%%%%%%%%%%%%%%%%%%%%%%%%%%%%%%%%%%%
\section*{References}
%%%%%%%%%%%%%%%%%%%%%%%%%%%%%%%%%%%%%%%%%%%%%%%%%%%%%%%%%%%%%%%

%%%%%%%%%%%%%%%%%%%%%%%%%%%%%%%%%%%%%%%%%%%%%%%%%%%%%%%%%%%%%%
\bibliography{sci,master}

\begin{thebibliography}{65}%
\makeatletter
\providecommand \@ifxundefined [1]{%
 \@ifx{#1\undefined}
}%
\providecommand \@ifnum [1]{%
 \ifnum #1\expandafter \@firstoftwo
 \else \expandafter \@secondoftwo
 \fi
}%
\providecommand \@ifx [1]{%
 \ifx #1\expandafter \@firstoftwo
 \else \expandafter \@secondoftwo
 \fi
}%
\providecommand \natexlab [1]{#1}%
\providecommand \enquote  [1]{``#1''}%
\providecommand \bibnamefont  [1]{#1}%
\providecommand \bibfnamefont [1]{#1}%
\providecommand \citenamefont [1]{#1}%
\providecommand \href@noop [0]{\@secondoftwo}%
\providecommand \href [0]{\begingroup \@sanitize@url \@href}%
\providecommand \@href[1]{\@@startlink{#1}\@@href}%
\providecommand \@@href[1]{\endgroup#1\@@endlink}%
\providecommand \@sanitize@url [0]{\catcode `\\12\catcode `\$12\catcode
  `\&12\catcode `\#12\catcode `\^12\catcode `\_12\catcode `\%12\relax}%
\providecommand \@@startlink[1]{}%
\providecommand \@@endlink[0]{}%
\providecommand \url  [0]{\begingroup\@sanitize@url \@url }%
\providecommand \@url [1]{\endgroup\@href {#1}{\urlprefix }}%
\providecommand \urlprefix  [0]{URL }%
\providecommand \Eprint [0]{\href }%
\providecommand \doibase [0]{http://dx.doi.org/}%
\providecommand \selectlanguage [0]{\@gobble}%
\providecommand \bibinfo  [0]{\@secondoftwo}%
\providecommand \bibfield  [0]{\@secondoftwo}%
\providecommand \translation [1]{[#1]}%
\providecommand \BibitemOpen [0]{}%
\providecommand \bibitemStop [0]{}%
\providecommand \bibitemNoStop [0]{.\EOS\space}%
\providecommand \EOS [0]{\spacefactor3000\relax}%
\providecommand \BibitemShut  [1]{\csname bibitem#1\endcsname}%
\let\auto@bib@innerbib\@empty
%</preamble>
\bibitem [{\citenamefont {Bender}\ and\ \citenamefont
  {Davidson}(1969)}]{Bender1969}%
  \BibitemOpen
  \bibfield  {author} {\bibinfo {author} {\bibfnamefont {C.~F.}\ \bibnamefont
  {Bender}}\ and\ \bibinfo {author} {\bibfnamefont {E.~R.}\ \bibnamefont
  {Davidson}},\ }\href {\doibase 10.1103/physrev.183.23} {\bibfield  {journal}
  {\bibinfo  {journal} {Phys. Rev.}\ }\textbf {\bibinfo {volume} {183}},\
  \bibinfo {pages} {23} (\bibinfo {year} {1969})}\BibitemShut {NoStop}%
\bibitem [{\citenamefont {Whitten}\ and\ \citenamefont
  {Hackmeyer}(1969)}]{Whitten1969}%
  \BibitemOpen
  \bibfield  {author} {\bibinfo {author} {\bibfnamefont {J.~L.}\ \bibnamefont
  {Whitten}}\ and\ \bibinfo {author} {\bibfnamefont {M.}~\bibnamefont
  {Hackmeyer}},\ }\href {\doibase 10.1063/1.1671985} {\bibfield  {journal}
  {\bibinfo  {journal} {J. Chem. Phys.}\ }\textbf {\bibinfo {volume} {51}},\
  \bibinfo {pages} {5584} (\bibinfo {year} {1969})}\BibitemShut {NoStop}%
\bibitem [{\citenamefont {Huron}, \citenamefont {Malrieu},\ and\ \citenamefont
  {Rancurel}(1973)}]{Huron1973}%
  \BibitemOpen
  \bibfield  {author} {\bibinfo {author} {\bibfnamefont {B.}~\bibnamefont
  {Huron}}, \bibinfo {author} {\bibfnamefont {J.~P.}\ \bibnamefont {Malrieu}},
  \ and\ \bibinfo {author} {\bibfnamefont {P.}~\bibnamefont {Rancurel}},\
  }\href {\doibase 10.1063/1.1679199} {\bibfield  {journal} {\bibinfo
  {journal} {J. Chem. Phys.}\ }\textbf {\bibinfo {volume} {58}},\ \bibinfo
  {pages} {5745} (\bibinfo {year} {1973})}\BibitemShut {NoStop}%
\bibitem [{\citenamefont {Buenker}\ and\ \citenamefont
  {Peyerimhoff}(1974)}]{Buenker1974}%
  \BibitemOpen
  \bibfield  {author} {\bibinfo {author} {\bibfnamefont {R.~J.}\ \bibnamefont
  {Buenker}}\ and\ \bibinfo {author} {\bibfnamefont {S.~D.}\ \bibnamefont
  {Peyerimhoff}},\ }\href {\doibase 10.1007/BF02394557} {\bibfield  {journal}
  {\bibinfo  {journal} {Theor. Chim. Acta}\ }\textbf {\bibinfo {volume} {35}},\
  \bibinfo {pages} {33} (\bibinfo {year} {1974})}\BibitemShut {NoStop}%
\bibitem [{\citenamefont {White}(1992)}]{White1992}%
  \BibitemOpen
  \bibfield  {author} {\bibinfo {author} {\bibfnamefont {S.~R.}\ \bibnamefont
  {White}},\ }\href {\doibase 10.1103/PhysRevLett.69.2863} {\bibfield
  {journal} {\bibinfo  {journal} {Phys. Rev. Lett.}\ }\textbf {\bibinfo
  {volume} {69}},\ \bibinfo {pages} {2863} (\bibinfo {year}
  {1992})}\BibitemShut {NoStop}%
\bibitem [{\citenamefont {White}(1993)}]{White1993}%
  \BibitemOpen
  \bibfield  {author} {\bibinfo {author} {\bibfnamefont {S.~R.}\ \bibnamefont
  {White}},\ }\href {\doibase 10.1103/PhysRevB.48.10345} {\bibfield  {journal}
  {\bibinfo  {journal} {Phys. Rev. B}\ }\textbf {\bibinfo {volume} {48}},\
  \bibinfo {pages} {10345} (\bibinfo {year} {1993})}\BibitemShut {NoStop}%
\bibitem [{\citenamefont {Chan}\ and\ \citenamefont {Sharma}(2011)}]{Chan2011}%
  \BibitemOpen
  \bibfield  {author} {\bibinfo {author} {\bibfnamefont {G.~K.-L.}\
  \bibnamefont {Chan}}\ and\ \bibinfo {author} {\bibfnamefont {S.}~\bibnamefont
  {Sharma}},\ }\href {\doibase 10.1146/annurev-physchem-032210-103338}
  {\bibfield  {journal} {\bibinfo  {journal} {Annu. Rev. Phys. Chem.}\ }\textbf
  {\bibinfo {volume} {62}},\ \bibinfo {pages} {465} (\bibinfo {year}
  {2011})}\BibitemShut {NoStop}%
\bibitem [{\citenamefont {Eriksen}, \citenamefont {Lipparini},\ and\
  \citenamefont {Gauss}(2017)}]{Eriksen2017}%
  \BibitemOpen
  \bibfield  {author} {\bibinfo {author} {\bibfnamefont {J.~J.}\ \bibnamefont
  {Eriksen}}, \bibinfo {author} {\bibfnamefont {F.}~\bibnamefont {Lipparini}},
  \ and\ \bibinfo {author} {\bibfnamefont {J.}~\bibnamefont {Gauss}},\ }\href
  {\doibase 10.1021/acs.jpclett.7b02075} {\bibfield  {journal} {\bibinfo
  {journal} {J. Phys. Chem. Lett.}\ }\textbf {\bibinfo {volume} {8}},\ \bibinfo
  {pages} {4633} (\bibinfo {year} {2017})}\BibitemShut {NoStop}%
\bibitem [{\citenamefont {Eriksen}\ and\ \citenamefont
  {Gauss}(2018)}]{Eriksen2018}%
  \BibitemOpen
  \bibfield  {author} {\bibinfo {author} {\bibfnamefont {J.~J.}\ \bibnamefont
  {Eriksen}}\ and\ \bibinfo {author} {\bibfnamefont {J.}~\bibnamefont
  {Gauss}},\ }\href {\doibase 10.1021/acs.jctc.8b00680} {\bibfield  {journal}
  {\bibinfo  {journal} {J. Chem. Theory Comput.}\ }\textbf {\bibinfo {volume}
  {14}},\ \bibinfo {pages} {5180} (\bibinfo {year} {2018})}\BibitemShut
  {NoStop}%
\bibitem [{\citenamefont {Eriksen}\ and\ \citenamefont
  {Gauss}(2019{\natexlab{a}})}]{Eriksen2019a}%
  \BibitemOpen
  \bibfield  {author} {\bibinfo {author} {\bibfnamefont {J.~J.}\ \bibnamefont
  {Eriksen}}\ and\ \bibinfo {author} {\bibfnamefont {J.}~\bibnamefont
  {Gauss}},\ }\href {\doibase 10.1021/acs.jctc.9b00456} {\bibfield  {journal}
  {\bibinfo  {journal} {J. Chem. Theory Comput.}\ }\textbf {\bibinfo {volume}
  {15}},\ \bibinfo {pages} {4873} (\bibinfo {year}
  {2019}{\natexlab{a}})}\BibitemShut {NoStop}%
\bibitem [{\citenamefont {Eriksen}\ and\ \citenamefont
  {Gauss}(2019{\natexlab{b}})}]{Eriksen2019b}%
  \BibitemOpen
  \bibfield  {author} {\bibinfo {author} {\bibfnamefont {J.~J.}\ \bibnamefont
  {Eriksen}}\ and\ \bibinfo {author} {\bibfnamefont {J.}~\bibnamefont
  {Gauss}},\ }\href {\doibase 10.1021/acs.jpclett.9b02968} {\bibfield
  {journal} {\bibinfo  {journal} {J. Phys. Chem. Lett.}\ }\textbf {\bibinfo
  {volume} {27}},\ \bibinfo {pages} {7910} (\bibinfo {year}
  {2019}{\natexlab{b}})}\BibitemShut {NoStop}%
\bibitem [{\citenamefont {Motta}\ and\ \citenamefont
  {Zhang}(2018)}]{Motta2018}%
  \BibitemOpen
  \bibfield  {author} {\bibinfo {author} {\bibfnamefont {M.}~\bibnamefont
  {Motta}}\ and\ \bibinfo {author} {\bibfnamefont {S.}~\bibnamefont {Zhang}},\
  }\href {\doibase 10.1002/wcms.1364} {\bibfield  {journal} {\bibinfo
  {journal} {WIREs Comput. Mol. Sci.}\ }\textbf {\bibinfo {volume} {8}},\
  \bibinfo {pages} {e1364} (\bibinfo {year} {2018})}\BibitemShut {NoStop}%
\bibitem [{\citenamefont {Lee}, \citenamefont {Malone},\ and\ \citenamefont
  {Reichman}(2020)}]{Lee2020}%
  \BibitemOpen
  \bibfield  {author} {\bibinfo {author} {\bibfnamefont {J.}~\bibnamefont
  {Lee}}, \bibinfo {author} {\bibfnamefont {F.~D.}\ \bibnamefont {Malone}}, \
  and\ \bibinfo {author} {\bibfnamefont {D.~R.}\ \bibnamefont {Reichman}},\
  }\href {\doibase 10.1063/5.0024835} {\bibfield  {journal} {\bibinfo
  {journal} {J. Chem. Phys.}\ }\textbf {\bibinfo {volume} {153}},\ \bibinfo
  {pages} {126101} (\bibinfo {year} {2020})}\BibitemShut {NoStop}%
\bibitem [{\citenamefont {Xu}, \citenamefont {Uejima},\ and\ \citenamefont
  {Ten-no}(2018)}]{Xu2018}%
  \BibitemOpen
  \bibfield  {author} {\bibinfo {author} {\bibfnamefont {E.}~\bibnamefont
  {Xu}}, \bibinfo {author} {\bibfnamefont {M.}~\bibnamefont {Uejima}}, \ and\
  \bibinfo {author} {\bibfnamefont {S.~L.}\ \bibnamefont {Ten-no}},\ }\href
  {\doibase 10.1103/PhysRevLett.121.113001} {\bibfield  {journal} {\bibinfo
  {journal} {Phys. Rev. Lett.}\ }\textbf {\bibinfo {volume} {121}},\ \bibinfo
  {pages} {113001} (\bibinfo {year} {2018})}\BibitemShut {NoStop}%
\bibitem [{\citenamefont {Xu}, \citenamefont {Uejima},\ and\ \citenamefont
  {Ten-no}(2020)}]{Xu2020}%
  \BibitemOpen
  \bibfield  {author} {\bibinfo {author} {\bibfnamefont {E.}~\bibnamefont
  {Xu}}, \bibinfo {author} {\bibfnamefont {M.}~\bibnamefont {Uejima}}, \ and\
  \bibinfo {author} {\bibfnamefont {S.~L.}\ \bibnamefont {Ten-no}},\ }\href
  {\doibase 10.1021/acs.jpclett.0c03084} {\bibfield  {journal} {\bibinfo
  {journal} {J. Phys. Chem. Lett.}\ }\textbf {\bibinfo {volume} {11}},\
  \bibinfo {pages} {9775} (\bibinfo {year} {2020})}\BibitemShut {NoStop}%
\bibitem [{\citenamefont {Magoulas}\ \emph {et~al.}(2021)\citenamefont
  {Magoulas}, \citenamefont {Gururangan}, \citenamefont {Piecuch},
  \citenamefont {Deustua},\ and\ \citenamefont {Shen}}]{Magoulas2021}%
  \BibitemOpen
  \bibfield  {author} {\bibinfo {author} {\bibfnamefont {I.}~\bibnamefont
  {Magoulas}}, \bibinfo {author} {\bibfnamefont {K.}~\bibnamefont
  {Gururangan}}, \bibinfo {author} {\bibfnamefont {P.}~\bibnamefont {Piecuch}},
  \bibinfo {author} {\bibfnamefont {J.~E.}\ \bibnamefont {Deustua}}, \ and\
  \bibinfo {author} {\bibfnamefont {J.}~\bibnamefont {Shen}},\ }\href {\doibase
  10.1021/acs.jctc.1c00181} {\bibfield  {journal} {\bibinfo  {journal} {J.
  Chem. Theory Comput.}\ }\textbf {\bibinfo {volume} {17}},\ \bibinfo {pages}
  {4006} (\bibinfo {year} {2021})}\BibitemShut {NoStop}%
\bibitem [{\citenamefont {Karthik}\ \emph {et~al.}(2021)\citenamefont
  {Karthik}, \citenamefont {J.~Emiliano}, \citenamefont {Jun},\ and\
  \citenamefont {Piotr}}]{Gururangan2021}%
  \BibitemOpen
  \bibfield  {author} {\bibinfo {author} {\bibfnamefont {G.}~\bibnamefont
  {Karthik}}, \bibinfo {author} {\bibfnamefont {D.}~\bibnamefont
  {J.~Emiliano}}, \bibinfo {author} {\bibfnamefont {S.}~\bibnamefont {Jun}}, \
  and\ \bibinfo {author} {\bibfnamefont {P.}~\bibnamefont {Piotr}},\ }\href
  {\doibase 10.1063/5.0064400} {\bibfield  {journal} {\bibinfo  {journal} {J.
  Chem. Phys.}\ }\textbf {\bibinfo {volume} {155}},\ \bibinfo {pages} {174114}
  (\bibinfo {year} {2021})}\BibitemShut {NoStop}%
\bibitem [{\citenamefont {Loos}, \citenamefont {Scemama},\ and\ \citenamefont
  {Jacquemin}(2020)}]{Loos2020a}%
  \BibitemOpen
  \bibfield  {author} {\bibinfo {author} {\bibfnamefont {P.-F.}\ \bibnamefont
  {Loos}}, \bibinfo {author} {\bibfnamefont {A.}~\bibnamefont {Scemama}}, \
  and\ \bibinfo {author} {\bibfnamefont {D.}~\bibnamefont {Jacquemin}},\ }\href
  {\doibase 10.1021/acs.jpclett.0c00014} {\bibfield  {journal} {\bibinfo
  {journal} {J. Phys. Chem. Lett.}\ }\textbf {\bibinfo {volume} {11}},\
  \bibinfo {pages} {2374} (\bibinfo {year} {2020})}\BibitemShut {NoStop}%
\bibitem [{\citenamefont {Eriksen}\ \emph {et~al.}(2020)\citenamefont
  {Eriksen}, \citenamefont {Anderson}, \citenamefont {Deustua}, \citenamefont
  {Ghanem}, \citenamefont {Hait}, \citenamefont {Hoffmann}, \citenamefont
  {Lee}, \citenamefont {Levine}, \citenamefont {Magoulas}, \citenamefont
  {Shen}, \citenamefont {Tubman}, \citenamefont {Whaley}, \citenamefont {Xu},
  \citenamefont {Yao}, \citenamefont {Zhang}, \citenamefont {Alavi},
  \citenamefont {Chan}, \citenamefont {Head-Gordon}, \citenamefont {Liu},
  \citenamefont {Piecuch}, \citenamefont {Sharma}, \citenamefont {Ten-no},
  \citenamefont {Umrigar},\ and\ \citenamefont {Gauss}}]{Eriksen2020}%
  \BibitemOpen
  \bibfield  {author} {\bibinfo {author} {\bibfnamefont {J.~J.}\ \bibnamefont
  {Eriksen}}, \bibinfo {author} {\bibfnamefont {T.~A.}\ \bibnamefont
  {Anderson}}, \bibinfo {author} {\bibfnamefont {J.~E.}\ \bibnamefont
  {Deustua}}, \bibinfo {author} {\bibfnamefont {K.}~\bibnamefont {Ghanem}},
  \bibinfo {author} {\bibfnamefont {D.}~\bibnamefont {Hait}}, \bibinfo {author}
  {\bibfnamefont {M.~R.}\ \bibnamefont {Hoffmann}}, \bibinfo {author}
  {\bibfnamefont {S.}~\bibnamefont {Lee}}, \bibinfo {author} {\bibfnamefont
  {D.~S.}\ \bibnamefont {Levine}}, \bibinfo {author} {\bibfnamefont
  {I.}~\bibnamefont {Magoulas}}, \bibinfo {author} {\bibfnamefont
  {J.}~\bibnamefont {Shen}}, \bibinfo {author} {\bibfnamefont {N.~M.}\
  \bibnamefont {Tubman}}, \bibinfo {author} {\bibfnamefont {K.~B.}\
  \bibnamefont {Whaley}}, \bibinfo {author} {\bibfnamefont {E.}~\bibnamefont
  {Xu}}, \bibinfo {author} {\bibfnamefont {Y.}~\bibnamefont {Yao}}, \bibinfo
  {author} {\bibfnamefont {N.}~\bibnamefont {Zhang}}, \bibinfo {author}
  {\bibfnamefont {A.}~\bibnamefont {Alavi}}, \bibinfo {author} {\bibfnamefont
  {G.~K.-L.}\ \bibnamefont {Chan}}, \bibinfo {author} {\bibfnamefont
  {M.}~\bibnamefont {Head-Gordon}}, \bibinfo {author} {\bibfnamefont
  {W.}~\bibnamefont {Liu}}, \bibinfo {author} {\bibfnamefont {P.}~\bibnamefont
  {Piecuch}}, \bibinfo {author} {\bibfnamefont {S.}~\bibnamefont {Sharma}},
  \bibinfo {author} {\bibfnamefont {S.~L.}\ \bibnamefont {Ten-no}}, \bibinfo
  {author} {\bibfnamefont {C.~J.}\ \bibnamefont {Umrigar}}, \ and\ \bibinfo
  {author} {\bibfnamefont {J.}~\bibnamefont {Gauss}},\ }\href {\doibase
  10.1021/acs.jpclett.0c02621} {\bibfield  {journal} {\bibinfo  {journal} {J.
  Phys. Chem. Lett.}\ }\textbf {\bibinfo {volume} {11}},\ \bibinfo {pages}
  {8922} (\bibinfo {year} {2020})}\BibitemShut {NoStop}%
\bibitem [{\citenamefont {Eriksen}(2021)}]{Eriksen2021}%
  \BibitemOpen
  \bibfield  {author} {\bibinfo {author} {\bibfnamefont {J.~J.}\ \bibnamefont
  {Eriksen}},\ }\href {\doibase 10.1021/acs.jpclett.0c03225} {\bibfield
  {journal} {\bibinfo  {journal} {J. Phys. Chem. Lett.}\ }\textbf {\bibinfo
  {volume} {12}},\ \bibinfo {pages} {418} (\bibinfo {year} {2021})}\BibitemShut
  {NoStop}%
\bibitem [{\citenamefont {Loos}, \citenamefont {Damour},\ and\ \citenamefont
  {Scemama}(2020)}]{Loos2020b}%
  \BibitemOpen
  \bibfield  {author} {\bibinfo {author} {\bibfnamefont {P.-F.}\ \bibnamefont
  {Loos}}, \bibinfo {author} {\bibfnamefont {Y.}~\bibnamefont {Damour}}, \ and\
  \bibinfo {author} {\bibfnamefont {A.}~\bibnamefont {Scemama}},\ }\href
  {\doibase 10.1063/5.0027617} {\bibfield  {journal} {\bibinfo  {journal} {J.
  Chem. Phys.}\ }\textbf {\bibinfo {volume} {153}},\ \bibinfo {pages} {176101}
  (\bibinfo {year} {2020})}\BibitemShut {NoStop}%
\bibitem [{\citenamefont {Caffarel}\ \emph
  {et~al.}(2016{\natexlab{a}})\citenamefont {Caffarel}, \citenamefont
  {Applencourt}, \citenamefont {Giner},\ and\ \citenamefont
  {Scemama}}]{Caffarel2016b}%
  \BibitemOpen
  \bibfield  {author} {\bibinfo {author} {\bibfnamefont {M.}~\bibnamefont
  {Caffarel}}, \bibinfo {author} {\bibfnamefont {T.}~\bibnamefont
  {Applencourt}}, \bibinfo {author} {\bibfnamefont {E.}~\bibnamefont {Giner}},
  \ and\ \bibinfo {author} {\bibfnamefont {A.}~\bibnamefont {Scemama}},\ }\href
  {\doibase 10.1063/1.4947093} {\bibfield  {journal} {\bibinfo  {journal} {J.
  Chem. Phys.}\ }\textbf {\bibinfo {volume} {144}},\ \bibinfo {pages} {151103}
  (\bibinfo {year} {2016}{\natexlab{a}})}\BibitemShut {NoStop}%
\bibitem [{\citenamefont {Holmes}, \citenamefont {Umrigar},\ and\ \citenamefont
  {Sharma}(2017)}]{Holmes2017}%
  \BibitemOpen
  \bibfield  {author} {\bibinfo {author} {\bibfnamefont {A.~A.}\ \bibnamefont
  {Holmes}}, \bibinfo {author} {\bibfnamefont {C.~J.}\ \bibnamefont {Umrigar}},
  \ and\ \bibinfo {author} {\bibfnamefont {S.}~\bibnamefont {Sharma}},\ }\href
  {\doibase 10.1063/1.4998614} {\bibfield  {journal} {\bibinfo  {journal} {J.
  Chem. Phys.}\ }\textbf {\bibinfo {volume} {147}},\ \bibinfo {pages} {164111}
  (\bibinfo {year} {2017})}\BibitemShut {NoStop}%
\bibitem [{\citenamefont {Chien}\ \emph {et~al.}(2018)\citenamefont {Chien},
  \citenamefont {Holmes}, \citenamefont {Otten}, \citenamefont {Umrigar},
  \citenamefont {Sharma},\ and\ \citenamefont {Zimmerman}}]{Chien2018}%
  \BibitemOpen
  \bibfield  {author} {\bibinfo {author} {\bibfnamefont {A.~D.}\ \bibnamefont
  {Chien}}, \bibinfo {author} {\bibfnamefont {A.~A.}\ \bibnamefont {Holmes}},
  \bibinfo {author} {\bibfnamefont {M.}~\bibnamefont {Otten}}, \bibinfo
  {author} {\bibfnamefont {C.~J.}\ \bibnamefont {Umrigar}}, \bibinfo {author}
  {\bibfnamefont {S.}~\bibnamefont {Sharma}}, \ and\ \bibinfo {author}
  {\bibfnamefont {P.~M.}\ \bibnamefont {Zimmerman}},\ }\href {\doibase
  10.1021/acs.jpca.8b01554} {\bibfield  {journal} {\bibinfo  {journal} {J.
  Phys. Chem. A}\ }\textbf {\bibinfo {volume} {122}},\ \bibinfo {pages} {2714}
  (\bibinfo {year} {2018})}\BibitemShut {NoStop}%
\bibitem [{\citenamefont {Loos}\ \emph {et~al.}(2018)\citenamefont {Loos},
  \citenamefont {Scemama}, \citenamefont {Blondel}, \citenamefont {Garniron},
  \citenamefont {Caffarel},\ and\ \citenamefont {Jacquemin}}]{Loos2018}%
  \BibitemOpen
  \bibfield  {author} {\bibinfo {author} {\bibfnamefont {P.~F.}\ \bibnamefont
  {Loos}}, \bibinfo {author} {\bibfnamefont {A.}~\bibnamefont {Scemama}},
  \bibinfo {author} {\bibfnamefont {A.}~\bibnamefont {Blondel}}, \bibinfo
  {author} {\bibfnamefont {Y.}~\bibnamefont {Garniron}}, \bibinfo {author}
  {\bibfnamefont {M.}~\bibnamefont {Caffarel}}, \ and\ \bibinfo {author}
  {\bibfnamefont {D.}~\bibnamefont {Jacquemin}},\ }\href {\doibase
  10.1021/acs.jctc.8b00406} {\bibfield  {journal} {\bibinfo  {journal} {J.
  Chem. Theory Comput.}\ }\textbf {\bibinfo {volume} {14}},\ \bibinfo {pages}
  {4360} (\bibinfo {year} {2018})}\BibitemShut {NoStop}%
\bibitem [{\citenamefont {Loos}\ \emph {et~al.}(2019)\citenamefont {Loos},
  \citenamefont {Boggio-Pasqua}, \citenamefont {Scemama}, \citenamefont
  {Caffarel},\ and\ \citenamefont {Jacquemin}}]{Loos2019}%
  \BibitemOpen
  \bibfield  {author} {\bibinfo {author} {\bibfnamefont {P.-F.}\ \bibnamefont
  {Loos}}, \bibinfo {author} {\bibfnamefont {M.}~\bibnamefont {Boggio-Pasqua}},
  \bibinfo {author} {\bibfnamefont {A.}~\bibnamefont {Scemama}}, \bibinfo
  {author} {\bibfnamefont {M.}~\bibnamefont {Caffarel}}, \ and\ \bibinfo
  {author} {\bibfnamefont {D.}~\bibnamefont {Jacquemin}},\ }\href {\doibase
  10.1021/acs.jctc.8b01205} {\bibfield  {journal} {\bibinfo  {journal} {J.
  Chem. Theory Comput.}\ }\textbf {\bibinfo {volume} {15}},\ \bibinfo {pages}
  {1939} (\bibinfo {year} {2019})}\BibitemShut {NoStop}%
\bibitem [{\citenamefont {Loos}\ \emph {et~al.}(2020)\citenamefont {Loos},
  \citenamefont {Lipparini}, \citenamefont {Boggio-Pasqua}, \citenamefont
  {Scemama},\ and\ \citenamefont {Jacquemin}}]{Loos2020c}%
  \BibitemOpen
  \bibfield  {author} {\bibinfo {author} {\bibfnamefont {P.~F.}\ \bibnamefont
  {Loos}}, \bibinfo {author} {\bibfnamefont {F.}~\bibnamefont {Lipparini}},
  \bibinfo {author} {\bibfnamefont {M.}~\bibnamefont {Boggio-Pasqua}}, \bibinfo
  {author} {\bibfnamefont {A.}~\bibnamefont {Scemama}}, \ and\ \bibinfo
  {author} {\bibfnamefont {D.}~\bibnamefont {Jacquemin}},\ }\href {\doibase
  10.1021/acs.jctc.9b01216} {\bibfield  {journal} {\bibinfo  {journal} {J.
  Chem. Theory Comput.}\ }\textbf {\bibinfo {volume} {16}},\ \bibinfo {pages}
  {1711} (\bibinfo {year} {2020})}\BibitemShut {NoStop}%
\bibitem [{\citenamefont {V{\'e}ril}\ \emph {et~al.}()\citenamefont
  {V{\'e}ril}, \citenamefont {Scemama}, \citenamefont {Caffarel}, \citenamefont
  {Lipparini}, \citenamefont {Boggio-Pasqua}, \citenamefont {Jacquemin},\ and\
  \citenamefont {Loos}}]{Veril2021}%
  \BibitemOpen
  \bibfield  {author} {\bibinfo {author} {\bibfnamefont {M.}~\bibnamefont
  {V{\'e}ril}}, \bibinfo {author} {\bibfnamefont {A.}~\bibnamefont {Scemama}},
  \bibinfo {author} {\bibfnamefont {M.}~\bibnamefont {Caffarel}}, \bibinfo
  {author} {\bibfnamefont {F.}~\bibnamefont {Lipparini}}, \bibinfo {author}
  {\bibfnamefont {M.}~\bibnamefont {Boggio-Pasqua}}, \bibinfo {author}
  {\bibfnamefont {D.}~\bibnamefont {Jacquemin}}, \ and\ \bibinfo {author}
  {\bibfnamefont {P.-F.}\ \bibnamefont {Loos}},\ }\href {\doibase
  https://doi.org/10.1002/wcms.1517} {\bibfield  {journal} {\bibinfo  {journal}
  {WIREs Comput. Mol. Sci.}\ }\textbf {\bibinfo {volume} {11}},\ \bibinfo
  {pages} {e1517}}\BibitemShut {NoStop}%
\bibitem [{\citenamefont {Giner}, \citenamefont {Scemama},\ and\ \citenamefont
  {Caffarel}(2013)}]{Giner2013}%
  \BibitemOpen
  \bibfield  {author} {\bibinfo {author} {\bibfnamefont {E.}~\bibnamefont
  {Giner}}, \bibinfo {author} {\bibfnamefont {A.}~\bibnamefont {Scemama}}, \
  and\ \bibinfo {author} {\bibfnamefont {M.}~\bibnamefont {Caffarel}},\ }\href
  {\doibase 10.1139/cjc-2013-0017} {\bibfield  {journal} {\bibinfo  {journal}
  {Can. J. Chem.}\ }\textbf {\bibinfo {volume} {91}},\ \bibinfo {pages} {879}
  (\bibinfo {year} {2013})}\BibitemShut {NoStop}%
\bibitem [{\citenamefont {Giner}, \citenamefont {Scemama},\ and\ \citenamefont
  {Caffarel}(2015)}]{Giner2015}%
  \BibitemOpen
  \bibfield  {author} {\bibinfo {author} {\bibfnamefont {E.}~\bibnamefont
  {Giner}}, \bibinfo {author} {\bibfnamefont {A.}~\bibnamefont {Scemama}}, \
  and\ \bibinfo {author} {\bibfnamefont {M.}~\bibnamefont {Caffarel}},\ }\href
  {\doibase 10.1063/1.4905528} {\bibfield  {journal} {\bibinfo  {journal} {J.
  Chem. Phys.}\ }\textbf {\bibinfo {volume} {142}},\ \bibinfo {pages} {044115}
  (\bibinfo {year} {2015})}\BibitemShut {NoStop}%
\bibitem [{\citenamefont {Caffarel}\ \emph
  {et~al.}(2016{\natexlab{b}})\citenamefont {Caffarel}, \citenamefont
  {Applencourt}, \citenamefont {Giner},\ and\ \citenamefont
  {Scemama}}]{Caffarel2016a}%
  \BibitemOpen
  \bibfield  {author} {\bibinfo {author} {\bibfnamefont {M.}~\bibnamefont
  {Caffarel}}, \bibinfo {author} {\bibfnamefont {T.}~\bibnamefont
  {Applencourt}}, \bibinfo {author} {\bibfnamefont {E.}~\bibnamefont {Giner}},
  \ and\ \bibinfo {author} {\bibfnamefont {A.}~\bibnamefont {Scemama}},\
  }\enquote {\bibinfo {title} {Using cipsi nodes in diffusion monte carlo},}\
  in\ \href {\doibase 10.1021/bk-2016-1234.ch002} {\emph {\bibinfo {booktitle}
  {Recent Progress in Quantum Monte Carlo}}}\ (\bibinfo {year} {2016})\
  Chap.~\bibinfo {chapter} {2}, pp.\ \bibinfo {pages} {15--46}\BibitemShut
  {NoStop}%
\bibitem [{\citenamefont {Garniron}\ \emph {et~al.}(2017)\citenamefont
  {Garniron}, \citenamefont {Scemama}, \citenamefont {Loos},\ and\
  \citenamefont {Caffarel}}]{Garniron2017}%
  \BibitemOpen
  \bibfield  {author} {\bibinfo {author} {\bibfnamefont {Y.}~\bibnamefont
  {Garniron}}, \bibinfo {author} {\bibfnamefont {A.}~\bibnamefont {Scemama}},
  \bibinfo {author} {\bibfnamefont {P.-F.}\ \bibnamefont {Loos}}, \ and\
  \bibinfo {author} {\bibfnamefont {M.}~\bibnamefont {Caffarel}},\ }\href
  {\doibase 10.1063/1.4992127} {\bibfield  {journal} {\bibinfo  {journal} {J.
  Chem. Phys.}\ }\textbf {\bibinfo {volume} {147}},\ \bibinfo {pages} {034101}
  (\bibinfo {year} {2017})}\BibitemShut {NoStop}%
\bibitem [{\citenamefont {Garniron}\ \emph {et~al.}(2018)\citenamefont
  {Garniron}, \citenamefont {Scemama}, \citenamefont {Giner}, \citenamefont
  {Caffarel},\ and\ \citenamefont {Loos}}]{Garniron2018}%
  \BibitemOpen
  \bibfield  {author} {\bibinfo {author} {\bibfnamefont {Y.}~\bibnamefont
  {Garniron}}, \bibinfo {author} {\bibfnamefont {A.}~\bibnamefont {Scemama}},
  \bibinfo {author} {\bibfnamefont {E.}~\bibnamefont {Giner}}, \bibinfo
  {author} {\bibfnamefont {M.}~\bibnamefont {Caffarel}}, \ and\ \bibinfo
  {author} {\bibfnamefont {P.~F.}\ \bibnamefont {Loos}},\ }\href {\doibase
  10.1063/1.5044503} {\bibfield  {journal} {\bibinfo  {journal} {J. Chem.
  Phys.}\ }\textbf {\bibinfo {volume} {149}},\ \bibinfo {pages} {064103}
  (\bibinfo {year} {2018})}\BibitemShut {NoStop}%
\bibitem [{\citenamefont {Garniron}\ \emph {et~al.}(2019)\citenamefont
  {Garniron}, \citenamefont {Gasperich}, \citenamefont {Applencourt},
  \citenamefont {Benali}, \citenamefont {Fert{\'e}}, \citenamefont {Paquier},
  \citenamefont {Pradines}, \citenamefont {Assaraf}, \citenamefont {Reinhardt},
  \citenamefont {Toulouse}, \citenamefont {Barbaresco}, \citenamefont {Renon},
  \citenamefont {David}, \citenamefont {Malrieu}, \citenamefont {V{\'e}ril},
  \citenamefont {Caffarel}, \citenamefont {Loos}, \citenamefont {Giner},\ and\
  \citenamefont {Scemama}}]{Garniron2019}%
  \BibitemOpen
  \bibfield  {author} {\bibinfo {author} {\bibfnamefont {Y.}~\bibnamefont
  {Garniron}}, \bibinfo {author} {\bibfnamefont {K.}~\bibnamefont {Gasperich}},
  \bibinfo {author} {\bibfnamefont {T.}~\bibnamefont {Applencourt}}, \bibinfo
  {author} {\bibfnamefont {A.}~\bibnamefont {Benali}}, \bibinfo {author}
  {\bibfnamefont {A.}~\bibnamefont {Fert{\'e}}}, \bibinfo {author}
  {\bibfnamefont {J.}~\bibnamefont {Paquier}}, \bibinfo {author} {\bibfnamefont
  {B.}~\bibnamefont {Pradines}}, \bibinfo {author} {\bibfnamefont
  {R.}~\bibnamefont {Assaraf}}, \bibinfo {author} {\bibfnamefont
  {P.}~\bibnamefont {Reinhardt}}, \bibinfo {author} {\bibfnamefont
  {J.}~\bibnamefont {Toulouse}}, \bibinfo {author} {\bibfnamefont
  {P.}~\bibnamefont {Barbaresco}}, \bibinfo {author} {\bibfnamefont
  {N.}~\bibnamefont {Renon}}, \bibinfo {author} {\bibfnamefont
  {G.}~\bibnamefont {David}}, \bibinfo {author} {\bibfnamefont {J.~P.}\
  \bibnamefont {Malrieu}}, \bibinfo {author} {\bibfnamefont {M.}~\bibnamefont
  {V{\'e}ril}}, \bibinfo {author} {\bibfnamefont {M.}~\bibnamefont {Caffarel}},
  \bibinfo {author} {\bibfnamefont {P.~F.}\ \bibnamefont {Loos}}, \bibinfo
  {author} {\bibfnamefont {E.}~\bibnamefont {Giner}}, \ and\ \bibinfo {author}
  {\bibfnamefont {A.}~\bibnamefont {Scemama}},\ }\href {\doibase
  10.1021/acs.jctc.9b00176} {\bibfield  {journal} {\bibinfo  {journal} {J.
  Chem. Theory Comput.}\ }\textbf {\bibinfo {volume} {15}},\ \bibinfo {pages}
  {3591} (\bibinfo {year} {2019})}\BibitemShut {NoStop}%
\bibitem [{\citenamefont {Damour}\ \emph {et~al.}(2021)\citenamefont {Damour},
  \citenamefont {V{\ifmmode\acute{e}\else\'{e}\fi}ril}, \citenamefont
  {Kossoski}, \citenamefont {Caffarel}, \citenamefont {Jacquemin},
  \citenamefont {Scemama},\ and\ \citenamefont {Loos}}]{Damour2021}%
  \BibitemOpen
  \bibfield  {author} {\bibinfo {author} {\bibfnamefont {Y.}~\bibnamefont
  {Damour}}, \bibinfo {author} {\bibfnamefont {M.}~\bibnamefont
  {V{\ifmmode\acute{e}\else\'{e}\fi}ril}}, \bibinfo {author} {\bibfnamefont
  {F.}~\bibnamefont {Kossoski}}, \bibinfo {author} {\bibfnamefont
  {M.}~\bibnamefont {Caffarel}}, \bibinfo {author} {\bibfnamefont
  {D.}~\bibnamefont {Jacquemin}}, \bibinfo {author} {\bibfnamefont
  {A.}~\bibnamefont {Scemama}}, \ and\ \bibinfo {author} {\bibfnamefont
  {P.-F.}\ \bibnamefont {Loos}},\ }\href {\doibase 10.1063/5.0065314}
  {\bibfield  {journal} {\bibinfo  {journal} {J. Chem. Phys.}\ }\textbf
  {\bibinfo {volume} {155}},\ \bibinfo {pages} {134104} (\bibinfo {year}
  {2021})}\BibitemShut {NoStop}%
\bibitem [{\citenamefont {Damour}\ \emph {et~al.}(2023)\citenamefont {Damour},
  \citenamefont {Quintero-Monsebaiz}, \citenamefont {Caffarel}, \citenamefont
  {Jacquemin}, \citenamefont {Kossoski}, \citenamefont {Scemama},\ and\
  \citenamefont {Loos}}]{Damour2023}%
  \BibitemOpen
  \bibfield  {author} {\bibinfo {author} {\bibfnamefont {Y.}~\bibnamefont
  {Damour}}, \bibinfo {author} {\bibfnamefont {R.}~\bibnamefont
  {Quintero-Monsebaiz}}, \bibinfo {author} {\bibfnamefont {M.}~\bibnamefont
  {Caffarel}}, \bibinfo {author} {\bibfnamefont {D.}~\bibnamefont {Jacquemin}},
  \bibinfo {author} {\bibfnamefont {F.}~\bibnamefont {Kossoski}}, \bibinfo
  {author} {\bibfnamefont {A.}~\bibnamefont {Scemama}}, \ and\ \bibinfo
  {author} {\bibfnamefont {P.-F.}\ \bibnamefont {Loos}},\ }\href {\doibase
  10.1021/acs.jctc.2c01111} {\bibfield  {journal} {\bibinfo  {journal} {J.
  Chem. Theory Comput.}\ }\textbf {\bibinfo {volume} {19}},\ \bibinfo {pages}
  {221} (\bibinfo {year} {2023})}\BibitemShut {NoStop}%
\bibitem [{\citenamefont {Schriber}\ and\ \citenamefont
  {Evangelista}(2016)}]{Schriber2016}%
  \BibitemOpen
  \bibfield  {author} {\bibinfo {author} {\bibfnamefont {J.~B.}\ \bibnamefont
  {Schriber}}\ and\ \bibinfo {author} {\bibfnamefont {F.~A.}\ \bibnamefont
  {Evangelista}},\ }\href {\doibase 10.1063/1.4948308} {\bibfield  {journal}
  {\bibinfo  {journal} {J. Chem. Phys.}\ }\textbf {\bibinfo {volume} {144}},\
  \bibinfo {pages} {161106} (\bibinfo {year} {2016})}\BibitemShut {NoStop}%
\bibitem [{\citenamefont {Tubman}\ \emph {et~al.}(2016)\citenamefont {Tubman},
  \citenamefont {Lee}, \citenamefont {Takeshita}, \citenamefont {Head-Gordon},\
  and\ \citenamefont {Whaley}}]{Tubman2016}%
  \BibitemOpen
  \bibfield  {author} {\bibinfo {author} {\bibfnamefont {N.~M.}\ \bibnamefont
  {Tubman}}, \bibinfo {author} {\bibfnamefont {J.}~\bibnamefont {Lee}},
  \bibinfo {author} {\bibfnamefont {T.~Y.}\ \bibnamefont {Takeshita}}, \bibinfo
  {author} {\bibfnamefont {M.}~\bibnamefont {Head-Gordon}}, \ and\ \bibinfo
  {author} {\bibfnamefont {K.~B.}\ \bibnamefont {Whaley}},\ }\href {\doibase
  10.1063/1.4955109} {\bibfield  {journal} {\bibinfo  {journal} {J. Chem.
  Phys.}\ }\textbf {\bibinfo {volume} {145}},\ \bibinfo {pages} {044112}
  (\bibinfo {year} {2016})}\BibitemShut {NoStop}%
\bibitem [{\citenamefont {Tubman}\ \emph {et~al.}(2018)\citenamefont {Tubman},
  \citenamefont {Levine}, \citenamefont {Hait}, \citenamefont {Head-Gordon},\
  and\ \citenamefont {Whaley}}]{Tubman2018}%
  \BibitemOpen
  \bibfield  {author} {\bibinfo {author} {\bibfnamefont {N.~M.}\ \bibnamefont
  {Tubman}}, \bibinfo {author} {\bibfnamefont {D.~S.}\ \bibnamefont {Levine}},
  \bibinfo {author} {\bibfnamefont {D.}~\bibnamefont {Hait}}, \bibinfo {author}
  {\bibfnamefont {M.}~\bibnamefont {Head-Gordon}}, \ and\ \bibinfo {author}
  {\bibfnamefont {K.~B.}\ \bibnamefont {Whaley}},\ }\href {\doibase
  10.48550/arXiv.1808.02049} {\bibfield  {journal} {\bibinfo  {journal}
  {arXiv}\ } (\bibinfo {year} {2018}),\ 10.48550/arXiv.1808.02049},\ \Eprint
  {http://arxiv.org/abs/1808.02049} {1808.02049} \BibitemShut {NoStop}%
\bibitem [{\citenamefont {Tubman}\ \emph {et~al.}(2020)\citenamefont {Tubman},
  \citenamefont {Freeman}, \citenamefont {Levine}, \citenamefont {Hait},
  \citenamefont {Head-Gordon},\ and\ \citenamefont {Whaley}}]{Tubman2020}%
  \BibitemOpen
  \bibfield  {author} {\bibinfo {author} {\bibfnamefont {N.~M.}\ \bibnamefont
  {Tubman}}, \bibinfo {author} {\bibfnamefont {C.~D.}\ \bibnamefont {Freeman}},
  \bibinfo {author} {\bibfnamefont {D.~S.}\ \bibnamefont {Levine}}, \bibinfo
  {author} {\bibfnamefont {D.}~\bibnamefont {Hait}}, \bibinfo {author}
  {\bibfnamefont {M.}~\bibnamefont {Head-Gordon}}, \ and\ \bibinfo {author}
  {\bibfnamefont {K.~B.}\ \bibnamefont {Whaley}},\ }\href {\doibase
  10.1021/acs.jctc.8b00536} {\bibfield  {journal} {\bibinfo  {journal} {J.
  Chem. Theory Comput.}\ }\textbf {\bibinfo {volume} {16}},\ \bibinfo {pages}
  {2139} (\bibinfo {year} {2020})}\BibitemShut {NoStop}%
\bibitem [{\citenamefont {Holmes}, \citenamefont {Tubman},\ and\ \citenamefont
  {Umrigar}(2016)}]{Holmes2016}%
  \BibitemOpen
  \bibfield  {author} {\bibinfo {author} {\bibfnamefont {A.~A.}\ \bibnamefont
  {Holmes}}, \bibinfo {author} {\bibfnamefont {N.~M.}\ \bibnamefont {Tubman}},
  \ and\ \bibinfo {author} {\bibfnamefont {C.~J.}\ \bibnamefont {Umrigar}},\
  }\href {\doibase 10.1021/acs.jctc.6b00407} {\bibfield  {journal} {\bibinfo
  {journal} {J. Chem. Theory Comput.}\ }\textbf {\bibinfo {volume} {12}},\
  \bibinfo {pages} {3674} (\bibinfo {year} {2016})}\BibitemShut {NoStop}%
\bibitem [{\citenamefont {Sharma}\ \emph {et~al.}(2017)\citenamefont {Sharma},
  \citenamefont {Holmes}, \citenamefont {Jeanmairet}, \citenamefont {Alavi},\
  and\ \citenamefont {Umrigar}}]{Sharma2017}%
  \BibitemOpen
  \bibfield  {author} {\bibinfo {author} {\bibfnamefont {S.}~\bibnamefont
  {Sharma}}, \bibinfo {author} {\bibfnamefont {A.~A.}\ \bibnamefont {Holmes}},
  \bibinfo {author} {\bibfnamefont {G.}~\bibnamefont {Jeanmairet}}, \bibinfo
  {author} {\bibfnamefont {A.}~\bibnamefont {Alavi}}, \ and\ \bibinfo {author}
  {\bibfnamefont {C.~J.}\ \bibnamefont {Umrigar}},\ }\href {\doibase
  10.1021/acs.jctc.6b01028} {\bibfield  {journal} {\bibinfo  {journal} {J.
  Chem. Theory Comput.}\ }\textbf {\bibinfo {volume} {13}},\ \bibinfo {pages}
  {1595} (\bibinfo {year} {2017})}\BibitemShut {NoStop}%
\bibitem [{\citenamefont {Yao}\ \emph {et~al.}(2020)\citenamefont {Yao},
  \citenamefont {Giner}, \citenamefont {Li}, \citenamefont {Toulouse},\ and\
  \citenamefont {Umrigar}}]{Yao2020}%
  \BibitemOpen
  \bibfield  {author} {\bibinfo {author} {\bibfnamefont {Y.}~\bibnamefont
  {Yao}}, \bibinfo {author} {\bibfnamefont {E.}~\bibnamefont {Giner}}, \bibinfo
  {author} {\bibfnamefont {J.}~\bibnamefont {Li}}, \bibinfo {author}
  {\bibfnamefont {J.}~\bibnamefont {Toulouse}}, \ and\ \bibinfo {author}
  {\bibfnamefont {C.~J.}\ \bibnamefont {Umrigar}},\ }\href {\doibase
  10.1063/5.0018577} {\bibfield  {journal} {\bibinfo  {journal} {J. Chem.
  Phys.}\ }\textbf {\bibinfo {volume} {153}},\ \bibinfo {pages} {124117}
  (\bibinfo {year} {2020})}\BibitemShut {NoStop}%
\bibitem [{\citenamefont {Yao}\ and\ \citenamefont {Umrigar}(2021)}]{Yao2021}%
  \BibitemOpen
  \bibfield  {author} {\bibinfo {author} {\bibfnamefont {Y.}~\bibnamefont
  {Yao}}\ and\ \bibinfo {author} {\bibfnamefont {C.~J.}\ \bibnamefont
  {Umrigar}},\ }\href {\doibase 10.1021/acs.jctc.1c00385} {\bibfield  {journal}
  {\bibinfo  {journal} {J. Chem. Theory Comput.}\ }\textbf {\bibinfo {volume}
  {17}},\ \bibinfo {pages} {4183} (\bibinfo {year} {2021})}\BibitemShut
  {NoStop}%
\bibitem [{\citenamefont {Larsson}\ \emph {et~al.}(2022)\citenamefont
  {Larsson}, \citenamefont {Zhai}, \citenamefont {Umrigar},\ and\ \citenamefont
  {Chan}}]{Larsson2022}%
  \BibitemOpen
  \bibfield  {author} {\bibinfo {author} {\bibfnamefont {H.~R.}\ \bibnamefont
  {Larsson}}, \bibinfo {author} {\bibfnamefont {H.}~\bibnamefont {Zhai}},
  \bibinfo {author} {\bibfnamefont {C.~J.}\ \bibnamefont {Umrigar}}, \ and\
  \bibinfo {author} {\bibfnamefont {G.~K.-L.}\ \bibnamefont {Chan}},\ }\href
  {\doibase 10.1021/jacs.2c06357} {\bibfield  {journal} {\bibinfo  {journal}
  {J. Am. Chem. Soc.}\ }\textbf {\bibinfo {volume} {144}},\ \bibinfo {pages}
  {15932} (\bibinfo {year} {2022})}\BibitemShut {NoStop}%
\bibitem [{\citenamefont {Liu}\ and\ \citenamefont {Hoffmann}(2014)}]{Liu2014}%
  \BibitemOpen
  \bibfield  {author} {\bibinfo {author} {\bibfnamefont {W.}~\bibnamefont
  {Liu}}\ and\ \bibinfo {author} {\bibfnamefont {M.}~\bibnamefont {Hoffmann}},\
  }\href {\doibase 10.1007/s00214-014-1481-x} {\bibfield  {journal} {\bibinfo
  {journal} {Theor. Chem. Acc.}\ }\textbf {\bibinfo {volume} {133}},\ \bibinfo
  {pages} {1481} (\bibinfo {year} {2014})}\BibitemShut {NoStop}%
\bibitem [{\citenamefont {Liu}\ and\ \citenamefont {Hoffmann}(2016)}]{Liu2016}%
  \BibitemOpen
  \bibfield  {author} {\bibinfo {author} {\bibfnamefont {W.}~\bibnamefont
  {Liu}}\ and\ \bibinfo {author} {\bibfnamefont {M.~R.}\ \bibnamefont
  {Hoffmann}},\ }\href {\doibase 10.1021/acs.jctc.5b01099} {\bibfield
  {journal} {\bibinfo  {journal} {J. Chem. Theory Comput.}\ }\textbf {\bibinfo
  {volume} {12}},\ \bibinfo {pages} {1169} (\bibinfo {year}
  {2016})}\BibitemShut {NoStop}%
\bibitem [{\citenamefont {Lei}, \citenamefont {Liu},\ and\ \citenamefont
  {Hoffmann}(2017)}]{Lei2017}%
  \BibitemOpen
  \bibfield  {author} {\bibinfo {author} {\bibfnamefont {Y.}~\bibnamefont
  {Lei}}, \bibinfo {author} {\bibfnamefont {W.}~\bibnamefont {Liu}}, \ and\
  \bibinfo {author} {\bibfnamefont {M.~R.}\ \bibnamefont {Hoffmann}},\ }\href
  {\doibase 10.1080/00268976.2017.1308029} {\bibfield  {journal} {\bibinfo
  {journal} {Mol. Phys.}\ }\textbf {\bibinfo {volume} {115}},\ \bibinfo {pages}
  {2696} (\bibinfo {year} {2017})}\BibitemShut {NoStop}%
\bibitem [{\citenamefont {Zhang}, \citenamefont {Liu},\ and\ \citenamefont
  {Hoffmann}(2020)}]{Zhang2020}%
  \BibitemOpen
  \bibfield  {author} {\bibinfo {author} {\bibfnamefont {N.}~\bibnamefont
  {Zhang}}, \bibinfo {author} {\bibfnamefont {W.}~\bibnamefont {Liu}}, \ and\
  \bibinfo {author} {\bibfnamefont {M.~R.}\ \bibnamefont {Hoffmann}},\ }\href
  {\doibase 10.1021/acs.jctc.9b01200} {\bibfield  {journal} {\bibinfo
  {journal} {J. Chem. Theory Comput.}\ }\textbf {\bibinfo {volume} {16}},\
  \bibinfo {pages} {2296} (\bibinfo {year} {2020})}\BibitemShut {NoStop}%
\bibitem [{\citenamefont {Zhang}, \citenamefont {Liu},\ and\ \citenamefont
  {Hoffmann}(2021)}]{Zhang2021}%
  \BibitemOpen
  \bibfield  {author} {\bibinfo {author} {\bibfnamefont {N.}~\bibnamefont
  {Zhang}}, \bibinfo {author} {\bibfnamefont {W.}~\bibnamefont {Liu}}, \ and\
  \bibinfo {author} {\bibfnamefont {M.~R.}\ \bibnamefont {Hoffmann}},\ }\href
  {\doibase 10.1021/acs.jctc.0c01187} {\bibfield  {journal} {\bibinfo
  {journal} {J. Chem. Theory Comput.}\ }\textbf {\bibinfo {volume} {17}},\
  \bibinfo {pages} {949} (\bibinfo {year} {2021})}\BibitemShut {NoStop}%
\bibitem [{\citenamefont {Coe}(2018)}]{Coe2018}%
  \BibitemOpen
  \bibfield  {author} {\bibinfo {author} {\bibfnamefont {J.~P.}\ \bibnamefont
  {Coe}},\ }\href {\doibase 10.1021/acs.jctc.8b00849} {\bibfield  {journal}
  {\bibinfo  {journal} {J. Chem. Theory Comput.}\ }\textbf {\bibinfo {volume}
  {14}},\ \bibinfo {pages} {5739} (\bibinfo {year} {2018})}\BibitemShut
  {NoStop}%
\bibitem [{\citenamefont {Coe}\ \emph {et~al.}(2022)\citenamefont {Coe},
  \citenamefont {Moreno~Carrascosa}, \citenamefont {Simmermacher},
  \citenamefont {Kirrander},\ and\ \citenamefont {Paterson}}]{Coe2022}%
  \BibitemOpen
  \bibfield  {author} {\bibinfo {author} {\bibfnamefont {J.~P.}\ \bibnamefont
  {Coe}}, \bibinfo {author} {\bibfnamefont {A.}~\bibnamefont
  {Moreno~Carrascosa}}, \bibinfo {author} {\bibfnamefont {M.}~\bibnamefont
  {Simmermacher}}, \bibinfo {author} {\bibfnamefont {A.}~\bibnamefont
  {Kirrander}}, \ and\ \bibinfo {author} {\bibfnamefont {M.~J.}\ \bibnamefont
  {Paterson}},\ }\href {\doibase 10.1021/acs.jctc.2c00738} {\bibfield
  {journal} {\bibinfo  {journal} {J. Chem. Theory Comput.}\ }\textbf {\bibinfo
  {volume} {18}},\ \bibinfo {pages} {6690} (\bibinfo {year}
  {2022})}\BibitemShut {NoStop}%
\bibitem [{\citenamefont {Booth}, \citenamefont {Thom},\ and\ \citenamefont
  {Alavi}(2009)}]{Booth2009}%
  \BibitemOpen
  \bibfield  {author} {\bibinfo {author} {\bibfnamefont {G.~H.}\ \bibnamefont
  {Booth}}, \bibinfo {author} {\bibfnamefont {A.~J.~W.}\ \bibnamefont {Thom}},
  \ and\ \bibinfo {author} {\bibfnamefont {A.}~\bibnamefont {Alavi}},\ }\href
  {\doibase 10.1063/1.3193710} {\bibfield  {journal} {\bibinfo  {journal} {J.
  Chem. Phys.}\ }\textbf {\bibinfo {volume} {131}},\ \bibinfo {pages} {054106}
  (\bibinfo {year} {2009})}\BibitemShut {NoStop}%
\bibitem [{\citenamefont {Cleland}, \citenamefont {Booth},\ and\ \citenamefont
  {Alavi}(2010)}]{Cleland2010}%
  \BibitemOpen
  \bibfield  {author} {\bibinfo {author} {\bibfnamefont {D.}~\bibnamefont
  {Cleland}}, \bibinfo {author} {\bibfnamefont {G.~H.}\ \bibnamefont {Booth}},
  \ and\ \bibinfo {author} {\bibfnamefont {A.}~\bibnamefont {Alavi}},\ }\href
  {\doibase 10.1063/1.3302277} {\bibfield  {journal} {\bibinfo  {journal} {J.
  Chem. Phys.}\ }\textbf {\bibinfo {volume} {132}},\ \bibinfo {pages} {041103}
  (\bibinfo {year} {2010})}\BibitemShut {NoStop}%
\bibitem [{\citenamefont {Blunt}\ \emph {et~al.}(2015)\citenamefont {Blunt},
  \citenamefont {Smart}, \citenamefont {Booth},\ and\ \citenamefont
  {Alavi}}]{Blunt2015}%
  \BibitemOpen
  \bibfield  {author} {\bibinfo {author} {\bibfnamefont {N.~S.}\ \bibnamefont
  {Blunt}}, \bibinfo {author} {\bibfnamefont {S.~D.}\ \bibnamefont {Smart}},
  \bibinfo {author} {\bibfnamefont {G.~H.}\ \bibnamefont {Booth}}, \ and\
  \bibinfo {author} {\bibfnamefont {A.}~\bibnamefont {Alavi}},\ }\href
  {\doibase 10.1063/1.4932595} {\bibfield  {journal} {\bibinfo  {journal} {J.
  Chem. Phys.}\ }\textbf {\bibinfo {volume} {143}},\ \bibinfo {pages} {134117}
  (\bibinfo {year} {2015})}\BibitemShut {NoStop}%
\bibitem [{\citenamefont {Ghanem}, \citenamefont {Lozovoi},\ and\ \citenamefont
  {Alavi}(2019)}]{Ghanem2019}%
  \BibitemOpen
  \bibfield  {author} {\bibinfo {author} {\bibfnamefont {K.}~\bibnamefont
  {Ghanem}}, \bibinfo {author} {\bibfnamefont {A.~Y.}\ \bibnamefont {Lozovoi}},
  \ and\ \bibinfo {author} {\bibfnamefont {A.}~\bibnamefont {Alavi}},\ }\href
  {\doibase 10.1063/1.5134006} {\bibfield  {journal} {\bibinfo  {journal} {J.
  Chem. Phys.}\ }\textbf {\bibinfo {volume} {151}},\ \bibinfo {pages} {224108}
  (\bibinfo {year} {2019})}\BibitemShut {NoStop}%
\bibitem [{\citenamefont {Deustua}, \citenamefont {Shen},\ and\ \citenamefont
  {Piecuch}(2017)}]{Deustua2017}%
  \BibitemOpen
  \bibfield  {author} {\bibinfo {author} {\bibfnamefont {J.~E.}\ \bibnamefont
  {Deustua}}, \bibinfo {author} {\bibfnamefont {J.}~\bibnamefont {Shen}}, \
  and\ \bibinfo {author} {\bibfnamefont {P.}~\bibnamefont {Piecuch}},\ }\href
  {\doibase 10.1103/PhysRevLett.119.223003} {\bibfield  {journal} {\bibinfo
  {journal} {Phys. Rev. Lett.}\ }\textbf {\bibinfo {volume} {119}},\ \bibinfo
  {pages} {223003} (\bibinfo {year} {2017})}\BibitemShut {NoStop}%
\bibitem [{\citenamefont {Deustua}\ \emph {et~al.}(2018)\citenamefont
  {Deustua}, \citenamefont {Magoulas}, \citenamefont {Shen},\ and\
  \citenamefont {Piecuch}}]{Deustua2018}%
  \BibitemOpen
  \bibfield  {author} {\bibinfo {author} {\bibfnamefont {J.~E.}\ \bibnamefont
  {Deustua}}, \bibinfo {author} {\bibfnamefont {I.}~\bibnamefont {Magoulas}},
  \bibinfo {author} {\bibfnamefont {J.}~\bibnamefont {Shen}}, \ and\ \bibinfo
  {author} {\bibfnamefont {P.}~\bibnamefont {Piecuch}},\ }\href {\doibase
  10.1063/1.5055769} {\bibfield  {journal} {\bibinfo  {journal} {J. Chem.
  Phys.}\ }\textbf {\bibinfo {volume} {149}},\ \bibinfo {pages} {151101}
  (\bibinfo {year} {2018})}\BibitemShut {NoStop}%
\bibitem [{\citenamefont {Burton}(2022)}]{Burton2021e}%
  \BibitemOpen
  \bibfield  {author} {\bibinfo {author} {\bibfnamefont {H.~G.~A.}\
  \bibnamefont {Burton}},\ }\href {\doibase 10.1021/acs.jctc.1c01089}
  {\bibfield  {journal} {\bibinfo  {journal} {J.\ Chem.\ Theory Comput.}\
  }\textbf {\bibinfo {volume} {18}},\ \bibinfo {pages} {1512} (\bibinfo {year}
  {2022})}\BibitemShut {NoStop}%
\bibitem [{\citenamefont {Marie}, \citenamefont {Burton},\ and\ \citenamefont
  {Loos}(2021)}]{Marie2021}%
  \BibitemOpen
  \bibfield  {author} {\bibinfo {author} {\bibfnamefont {A.}~\bibnamefont
  {Marie}}, \bibinfo {author} {\bibfnamefont {H.~G.~A.}\ \bibnamefont
  {Burton}}, \ and\ \bibinfo {author} {\bibfnamefont {P.-F.}\ \bibnamefont
  {Loos}},\ }\href {\doibase 10.1088/1361-648X/abe795} {\bibfield  {journal}
  {\bibinfo  {journal} {J.\ Phys.: Condens.\ Matter}\ }\textbf {\bibinfo
  {volume} {33}},\ \bibinfo {pages} {283001} (\bibinfo {year}
  {2021})}\BibitemShut {NoStop}%
\bibitem [{\citenamefont {Goodson}(2012)}]{Goodson2012}%
  \BibitemOpen
  \bibfield  {author} {\bibinfo {author} {\bibfnamefont {D.~Z.}\ \bibnamefont
  {Goodson}},\ }\href {\doibase 10.1002/wcms.92} {\bibfield  {journal}
  {\bibinfo  {journal} {WIREs Comput.\ Mol.\ Sci.}\ }\textbf {\bibinfo {volume}
  {2}},\ \bibinfo {pages} {743} (\bibinfo {year} {2012})}\BibitemShut {NoStop}%
\bibitem [{\citenamefont {Goodson}(2019)}]{Goodson2019}%
  \BibitemOpen
  \bibfield  {author} {\bibinfo {author} {\bibfnamefont {D.~Z.}\ \bibnamefont
  {Goodson}},\ }in\ \href {\doibase
  https://doi.org/10.1016/B978-0-12-813651-5.00009-7} {\emph {\bibinfo
  {booktitle} {Mathematical Physics in Theoretical Chemistry}}},\ \bibinfo
  {series and number} {{Developments in Physical {\&} Theoretical Chemistry}},\
  \bibinfo {editor} {edited by\ \bibinfo {editor} {\bibfnamefont
  {S.}~\bibnamefont {Blinder}}\ and\ \bibinfo {editor} {\bibfnamefont
  {J.}~\bibnamefont {House}}}\ (\bibinfo  {publisher} {Elsevier},\ \bibinfo
  {year} {2019})\ p.\ \bibinfo {pages} {295}\BibitemShut {NoStop}%
\bibitem [{\citenamefont {Mayer}\ and\ \citenamefont {Tong}(1985)}]{Mayer1985}%
  \BibitemOpen
  \bibfield  {author} {\bibinfo {author} {\bibfnamefont {I.~L.}\ \bibnamefont
  {Mayer}}\ and\ \bibinfo {author} {\bibfnamefont {B.~Y.}\ \bibnamefont
  {Tong}},\ }\href {\doibase 10.1088/0022-3719/18/17/008} {\ \textbf {\bibinfo
  {volume} {18}},\ \bibinfo {pages} {3297} (\bibinfo {year}
  {1985})}\BibitemShut {NoStop}%
\bibitem [{\citenamefont {L{\"{o}}wdin}(1951)}]{Lowdin1951}%
  \BibitemOpen
  \bibfield  {author} {\bibinfo {author} {\bibfnamefont {P.~O.}\ \bibnamefont
  {L{\"{o}}wdin}},\ }\href {\doibase 10.1063/1.1748067} {\bibfield  {journal}
  {\bibinfo  {journal} {J.\ Chem.\ Phys.}\ }\textbf {\bibinfo {volume} {19}},\
  \bibinfo {pages} {1396} (\bibinfo {year} {1951})}\BibitemShut {NoStop}%
\bibitem [{\citenamefont {Malrieu}, \citenamefont {Durand},\ and\ \citenamefont
  {Daudey}(1985)}]{Malrieu1985}%
  \BibitemOpen
  \bibfield  {author} {\bibinfo {author} {\bibfnamefont {J.~P.}\ \bibnamefont
  {Malrieu}}, \bibinfo {author} {\bibfnamefont {P.}~\bibnamefont {Durand}}, \
  and\ \bibinfo {author} {\bibfnamefont {J.~P.}\ \bibnamefont {Daudey}},\
  }\href {\doibase 10.1088/0305-4470/18/5/014} {\bibfield  {journal} {\bibinfo
  {journal} {J. Phys. A: Math. Theor.}\ }\textbf {\bibinfo {volume} {18}},\
  \bibinfo {pages} {809} (\bibinfo {year} {1985})}\BibitemShut {NoStop}%
\end{thebibliography}%
%%%%%%%%%%%%%%%%%%%%%%%%%%%%%%%%%%%%%%%%%%%%%%%%%%%%%%%%%%%%%%

\end{document}